\documentclass[a4paper,11pt]{article}
\usepackage{amsfonts,amsmath,amssymb,bm,url}
\usepackage{graphicx}
\usepackage{epstopdf}
\usepackage{xcolor}  
\graphicspath{{./CrustcoreB_figs/} }
\usepackage{lipsum}
\usepackage{float}
\usepackage{subfigure}
\usepackage{cleveref}

\title{Estimating magnetar radii with an empirical meta-model}

\author{Debarati Chatterjee\thanks{dchatterjee@lpccaen.in2p3.fr} $\>$  and Francesca Gulminelli  \\
	LPC, UMR6534, ENSICAEN, F-14050 Caen, France  \\
	\and 
	Debora P. Menezes \\
	UFSC, Dept de Fisica-CFM, Florianopolis, Brazil \\
	}

\date{\today}

\begin{document}

\maketitle

\begin{abstract}
The presence of strong magnetic fields in neutron stars, such as in magnetars, may significantly affect their crust-core transition properties and the crust size. This knowledge is crucial in the correct interpretation of astrophysical phenomena involving magnetars, such as glitches in observed rotation frequencies, cooling, bursts and possibly tidal polarizabilities.
A recently developed meta-modelling technique allows exploring the model dependence of density functional theory equation of state calculations.
In this work, we extend this metamodel to investigate the effect of strong magnetic fields on spinodal instabilities of neutron star matter and the associated  crust-core properties. Both Tolman-Oppenheimer-Volkov and a full self-consistent numerical calculations are performed for the neutron star structure, the results being quantitatively different for strong magnetic fields.
\end{abstract}

\section{Introduction}

\indent With the rapid advance of astrophysical techniques on compact objects observation, there is an increased need to improve theoretical modelling to accurately describe the associated observed phenomena. Following the recent launch of the NICER (Neutron Star Interior Composition Explorer) mission \cite{Gendereau, Ozel}, one expects to be able to measure neutron star radii to 5\% accuracy. One may also expect improved estimates of the moment of inertia of neutron stars from gravitational wave observations \cite{Bandyopadhyay}. However, improvements in theoretical estimates of neutron star radii or moment of inertia are limited due to the uncertainties in the properties of nuclear matter at high densities \cite{Fortin}. One such property is the density at which nuclear matter in neutron stars goes through a phase transition, from solid inhomogeneous matter in the crust to the homogeneous liquid phase in the core. Although the liquid-gas phase transition is well understood from the studies of homogeneous nuclear matter \cite{Gulminelli2013}, in the case of neutron star matter the problem is more complex.
\\

The understanding of the properties of the crust-core
interface is important for accurate interpretation of several other astrophysical 
phenomena involving neutron stars.  Explanations for glitches in the otherwise periodic rotational periods of neutron stars requires a knowledge of their elastic properties and crustal moment of inertia \cite{Fattoyev}. Thermal properties of the crust-core interface govern their cooling behavior \cite{Potekhin}. Viscosity of the crust-core transition region also plays a major role in the emission of gravitational waves due to non-axisymmetric instabilities \cite{Bondarescu}.
In 2017, the first detection of a gravitational wave, GW170817, generated by the collision of two neutron stars was confirmed by the LIGO and Virgo collaborations \cite{PRL119} and also observed in different regions of the electromagnetic spectrum, what was then called a multi-messenger event \cite{AJL848}. To reproduce GW170817 tidal polarizability \cite{Damour86, Flanagan87}, besides the use of an appropriate equation of state, the size of the crust itself plays a non negligible role in the calculation of the Love number, which is a necessary step to compute the tidal polarizabilities of the neutron stars belonging to a binary system \cite{love}. 
\\
In non-magnetized neutron stars, the outer crust is generally described by the BPS equation of state \cite{bps} and the inner crust can be described by the BBP equation of state \cite{bbp} or by a pasta phase region.
Different calculations and numerical techniques
can be used to investigate the crust-core region, the most common
involving binodal sections, thermodynamical and dynamical
spinodals \cite{Ducoin2010} and the pasta phase transition to
homogeneous phase \cite{pasta}, all of them converging to comparable
transition densities \cite{allcalc}.
The crust-core transition being an equilibrium phenomenon, a 
quantitative estimation of the transition point in principle demands a complete
pasta phase calculation.
 However, such modelling is complex and it relies on a proper
determination of the surface tension,  
 a matter of debate in the
literature, at least for the last 40 years \cite{ravenhall}. To circumvent
the pasta calculation, another possibility is to study the transition region
from the properties of the homogeneous matter and its phase transition
by determining the appearance of the instability region well described
by spinodals. In this case, many properties of the crust-core
interface and of the crust itself can be obtained directly from the transition
densities and related thermodynamical quantities. According to
well established calculations
\cite{lorenz93,ravenhall94,link99,lattimer2007,jorge2010}, many of the
crust properties, as its moment of inertia, mass and radius can be
directly computed from the equation of state at the transition
boundary. In particular, it was shown in ref. \cite{Zdunik},  that the approximate results for the crust extracted from the core calculation up to the transition density lead to very accurate results for both its 
thickness and mass.
\\
However the different microscopic (relativistic and non-relativistic) calculations give different predictions for the crust-core transition point \cite{Ducoin2010, Ducoin2008,Ducoin2011}. These differences may arise either due to the differences in the underlying form of the density functionals inherent to the model or due to limitations in our knowledge of the empirical nuclear parameters extracted from experiments\cite{Carreau,Antic}. 
Hence, the crust thickness is generally model dependent, but very important if one wants to study different astrophysical phenomena involving isolated or binaries of neutron stars.

Magnetars are a special class of isolated neutron stars
bearing surface magnetic fields three orders of magnitude larger than
their non-magnetized counterparts and slow rotation. Although just about 30 of them
have been clearly identified \cite{catalogue} so far,
they display a large number of peculiar phenomena, such as
antiglitches, bursts and oscillations that challenge our understanding of neutron stars.
It is well known that in the presence of strong
magnetic fields, the motion of charged particles are affected, which in turn, alters the equation of state (EoS) \cite{Chakrabarty}
and related properties both in the core and the crust.
Consequently the crust-core boundary of magnetars presents
different features as compared with the usual one of neutron stars. 
Moreover, recent studies \cite{pons2013} indicate that during the cooling process of
a magnetar, the crust-core transition region plays an important role
in determining its final configuration. 
Another interesting calculation \cite{jorge2014}
points to quite a large crust (2.4 Km) for a 1.4 $M_\odot$ neutron
star, with non-negligible consequences
on the cooling mechanism and on the emission of gravitational waves.
\\
The crust-core region of magnetars has recently been studied within the
relativistic mean field model framework \cite{cp1,cp2,cp3}. In these
works, the effects of strong magnetic fields ($10^{15} - 10^{17}$ G)
on the instability region have been investigated with the help of the
Vlasov formalism used to determine the dynamical spinodals.  
Although the magnetic fields at the surface of magnetars are of the
order of $10^{14} - 10^{15}$ G \cite{catalogue}, stronger fields can
be expected in their interior. Assuming that the crust can be as large as 
it was computed in \cite{jorge2014}, fields of the order of $10^{17}$ G
would be justified at the transition boundary.
\\

In the present work, we revisit the problem of the crust-core boundary
and the crust size in magnetars by studying thermodynamical and dynamical spinodal
sections within different models and formalisms from the ones used in
\cite{cp1,cp2,cp3}. 
We use a meta-modelling technique based on the density
functional theory, which consists of a general framework developed to take into
account a huge variety of models usually used to describe nuclear and 
stellar matter \cite{Casali1,Casali2,Chatterjee}. 
The parameters of this meta-model are directly related 
to the empirical parameters that can be constrained through nuclear 
physics experiments \cite{Casali1}. 
This allows us to determine the most influential parameters governing
the crust-core transition in the presence of strong magnetic fields, and to assess
the model dependence of the results. 
\\

The relation between the EoS, the transition point, and the crustal thickness is 
complicated by the fact that the Tolman Oppenheimer Volkov (TOV) formalism 
of hydrostatic equilibrium in general relativity is not correct in the presence of 
strong magnetic fields, since they break the spherical symmetry assumed in the calculations that lead to  the TOV \cite{Chatterjee2014}. 
In the present work, we employ a recently developed consistent formalism for constructing numerical models of neutron stars, considering magnetic field effects on the microscopic EoS and solving the Einstein-Maxwell and equilibrium equations to obtain the global neutron star structure. 
\\

This paper is organized as follows: in section \ref{sec:theo}, we describe the theoretical formalism for the microscopic calculations. In section \ref{sec:metamodel} we recapitulate the details of this MetaModel (MM) approach for the Equation of State (EoS), while the technique for the determination of the crust-core (CC) transition point is detailed in section \ref{sec:ccpt}. We study the effect of strong magnetic fields on the CC point in section \ref{sec:mag}, and we examine the sensitivity to the most influential empirical parameters in the same section. We show in particular (section \ref{sec:sensrho_dyn}) that the influence of the poorely constrained derivatives of the symmetry energy ($L_{sym}$, $K_{sym}$) dramatically increases with the increase of the magnetic field. 
The effect of magnetic fields on the neutron star structure is explored in Sec. \ref{sec:massradius}, and we show that going beyond the spherical TOV approximation is very important not only for the determination of the M(R) relation \cite{Chatterjee2014,Chatterjee2016,Chatterjee2018}, but also for the determination of the crustal properties even for relatively low magnetic fields. Finally we infer the conclusions of the study in section \ref{sec:conclusions}.

\section{Theoretical formalism for the microscopic model} 
\label{sec:theo}

In this section, we elaborate on the  theoretical approach employed to describe the microscopic neutron star matter.
In section \ref{sec:metamodel}, we highlight the fundamental aspects of the MM, that was originally proposed in \cite{Casali1}, and applied to study neutron stars \cite{Casali2,Carreau,Antic} as well as nuclei \cite{Chatterjee}. In the following Sec. \ref{sec:ccpt}, we apply this MM to study the phase transition properties and to determine the crust-core phase boundary.

\subsection{Meta-modelling approach for EoS}
\label{sec:metamodel}

In previous works on the effect of the magnetic field on the crust-core phase transition \cite{cp1,cp2,cp3}, EoS functionals extracted from the relativistic mean-field approach were employed. To have a complementary view on the subject, in this work we take a wide range of reference models, some the most widely used and realistic non-relativistic functionals, namely the Skyrme functional SLy5 and SLy4 \cite{Chabanat}, Bsk17 \cite{Bsk17} as well as the relativistic model TM1 \cite{TM1}. The choice was inspired by the fact that their values of slope of symmetry energy cover a wide range, which is known to influence crust-core transition properties \cite{Ducoin2010,Carreau}. \\ 

In order to explore these different models, we will use 
the technique proposed in ref. \cite{Casali1}, where a generic meta-modelling of the nuclear EoS was introduced. In that reference, a flexible fully analytical functional was proposed with a parameter space large enough to be able to accurately reproduce different EoS functional forms from phenomenological as well as ab-initio calculations. 
By changing the parameter set of the meta-functional, one is then able to switch between different models with a very limited computational effort. Moreover, since all the parameters can be independently varied, interpolations are possible between the different models and sensitivity studies can be done, in order to assess the relative influence of the parameters on a given observable. We will take advantage of this property of the meta-modelling in section \ref{sec:sensrho_dyn}.

The energy per particle of homogeneous nuclear matter at baryonic density $n=n_n+n_p$
and asymmetry $\delta=(n_n-n_p)/n$ is expressed as: 

\begin{eqnarray}
e_{HNM} (n,\delta) = 
e_{kin}^p +e_{kin}^n+ 
+ \sum_{\alpha=0}^4 (a_{\alpha is} + a_{\alpha iv} \delta^2) \frac{1}{\alpha!}\left (\frac{n-n_{sat}}{n_{sat}}\right)^\alpha u_{\alpha} (n,b) \,~.
\label{eq:e_hnm}
\end{eqnarray}

In this expression, $n_{sat}$ is the saturation density of symmetric matter,
$e_{kin_q}$, $q=n,p$ is the kinetic energy functional

\begin{equation}
e_{kin}^q = \frac{3\hbar^2}{10 m_q^*} \left( \frac{3 \pi^2 n}{2} \right)^{2/3} (1 + \tau_3\delta)^{2/3}  
\end{equation}

and the in-medium neutron and proton effective masses $m_n^*$ and $m_p^*$ are also expanded
in terms of the density parameter $x$ as~\cite{Casali1}:
\begin{equation}
\frac{m}{m^*_q} = 1+ \left( \kappa_{sat} + \tau_3 \kappa_{sym} \delta \right) \frac{n}{n_{sat}},
\label{eq:effm_expn}
\end{equation}
where $\tau_3=1$ for neutrons and $-1$ for protons.
The $a_{\alpha,is}$ and $a_{\alpha,iv}$ parameters fulfill a one-to-one mapping with the so-called empirical parameters, given by the successive derivatives around saturation $n_{sat}$ of the isoscalar ($E_{sat}$, $K_{sat}$, $Q_{sat}$, $Z_{sat}$) and isovector ($E_{sym}$, $L_{sym}$, $K_{sym}$, $Q_{sym}$, $Z_{sym}$)
part of the energy functional.
Finally, $u_{\alpha} (n,b)$ is a low-density correction that insures the correct zero density limit of the functional, and becomes negligible at a density $n_{min}\ll n_{sat}$. The value of $n_{min}$ is governed by the extra parameter $b$ of the model. 
\\

This meta-functional was fitted to a large number of relativistic and non-relativistic nuclear models, and the corresponding parameter sets 
$\{\vec {X}\}$ = $ \{ n_{sat}$, $E_{sat}$, $K_{sat}$, $Q_{sat}$, $Z_{sat}$, $E_{sym}$, $L_{sym}$, $K_{sym}$, $Q_{sym}$, $Z_{sym}$, $m^*/m$, $\Delta m^*/m \}$ are given in table X and XI of ref.\cite{Casali1}. 
The technique was shown to give a very accurate reproduction of the equation of state of catalyzed 
neutron star matter \cite{Casali1}. In that fit, the extra parameter $b$ of the meta-modelling 
was taken as a constant $b=10ln(2)\approx 6.93$. The reason of that choice was that this parameter,
governing the behavior of the energy functional at extremely low density,  plays a negligible role on the energy and pressure
of symmetric and pure neutron matter.
However, when looking at instability properties of $\beta$-equilibrium matter, the behavior at extremely low proton density, where the validity of the Taylor expansion around saturation breaks down, becomes crucial, and $b$ has to be included as an extra parameter. This parameter can be fitted to the spinodal curve when one wants to reproduce a specific existing model, or can be largely varied when one wants to explore the model dependence of the predictions.
\\

A fit of the $b$ parameter on the spinodal curve was performed for a few reference models, including SLy5 and SLy4, in \cite{Antic}. In addition to that,
in that work, a posterior distribution of the $b$ parameter was obtained within a complete Bayesian analysis, under the constraint that the different functionals described by the meta-modelling parameter space should all fall within the uncertainty interval obtained at low densities from the ab-initio effective field theory calculations of ref.\cite{Drischler}. 
The best reproduction of the original Skyrme models was found for $b = 10/3 log 2 = \sim 2.31$ so we will use this optimized value for the rest of this study.
\\

When dealing with inhomogeneous matter as in Section \ref{sec:ccpt} below, the meta-functional is extended by adding two extra gradient terms\cite{Chatterjee}:

\begin{eqnarray}
e(n,\delta) =
 e_{HNM}(n,\delta)  
+ \sum_{q=n,p} \frac{\hbar^2}{2m_q^*} \frac{ \tau_{2q}}{n} 
+  C_{fin} 
\frac{ \left( \boldsymbol{\nabla}n \right)^2}{n}
+  D_{fin} 
\frac{ \left( \boldsymbol{\nabla}(n\delta) \right)^2}{n}
. 
\label{eq_sym_density_energy_Skyrme}
\end{eqnarray}
  
The $\tau_{2q}=\tau_{2q}^l + \tau_{2q}^{nl}$ are second order local and non-local corrections
arising from an $\hbar$ expansion of the kinetic energy density of an inhomogeneous system:
\begin{eqnarray}
\tau_{2q}^l    & =& \frac{1}{36} \frac{ \left( \boldsymbol{\nabla}n_q \right)^2 }{n_q} + \frac{1}{3} \Delta n_q  \label{tau_loc}\\
\tau_{2q}^{nl} & =&
\frac{1}{6} \frac{ \boldsymbol{\nabla}n_q \boldsymbol{\nabla} f_q}{f_q}
+\frac{1}{6}  n_0  \frac{\Delta f_q}{f_q}
- \frac{1}{12}  n_q  \left(  \frac{\boldsymbol{\nabla} f_q}{f_q} \right) ^2, \label{tau_nloc}
\end{eqnarray}
with $f_q=m/m_q^*$. 
The $C_{fin}$ ($D_{fin}$) parameters govern the isoscalar (isovector) surface properties of the energy functional in the linear response approximation. Such terms appear explicitly in Skyrme functionals,  and their values are  determined in the fitting protocol of Skyrme parameters to nuclear ground state properties, together with the other parameters of the functional. In the case of the meta-modelling representation of non-Skyrme functionals, we determine the values of $C_{fin}$ and $D_{fin}$ through a $\chi^2$ fit of the masses of a set of magic and semi-magic nuclei, where the theoretical masses are calculated in the Extended-Thomas-Fermi spherical approximation with the meta-functional using a parametrized density profile \cite{Antic}.

  The resulting bulk and surface parameters are reported in table \ref{tab:models} and \ref{tab:surface} for the functionals used in this work.

\begin{table*}[htbp]
   \caption{Empirical parameters for reference models }
\begin{tabular}{|c|c|c|c|c|c|c|c|c|c|c|c|c|}
\hline
   Model & $n_{sat}$ & $E_{sat}$ & $K_{sat}$ & $Q_{sat}$ & $Z_{sat}$ & $E_{sym}$ & $L_{sym}$ & $K_{sym}$ & $Q_{sym}$ & $Z_{sym}$ & $m^*/m$ & $\Delta m^/m$ \\
 \hline
{SLy5} & 0.1604 & -15.98 & 230 & -364 & 1592 & 32.03 & 48.3 & -112 & 501 & -3087 & 0.70 & -0.18 \\
\hline
{SLy4} & 0.1595 & -15.97 & 230 & -363 & 1587 & 32.01 & 46.0 & -120 & 521 & -3197 & 0.69 & -0.19 \\
{TM1} &  0.1450 & -16.26 & 281 & -285 & 2014 & 36.94 & 111.0 & 34.0 & -67. & -1546.0 & 0.71 & -0.09 \\
\hline
{Bsk17} & 0.1586 & -16.05 & 242 & -364 & 1460 & 30.00 & 36.3 & -182 & 451 & -2508 & 0.80 & 0.04 \\
\hline
\end{tabular}
\label{tab:models}
\end{table*}

\begin{table*}[htbp]
   \caption{Surface parameters for reference models }
\begin{tabular}{|c|c|c|c|}
\hline
  Model & $ C_{fin} (MeV fm^5)$ & $D_{fin}(MeV fm^5)$\\
 \hline
 {SLy5} & 56.25 & 23.95 \\
\hline
 {SLy4} & 122.68 & 185.31 \\
 \hline
 {TM1} & 55.0 & 100. \\
 \hline
 {Bsk17} & 35.0 & 10. \\
 \hline
\end{tabular}
\label{tab:surface}
\end{table*}

\newpage
\subsection{Crust-core phase transition (CCPT) properties}
\label{sec:ccpt} 

\indent Once the meta-functional for the baryonic energy of section \ref{sec:metamodel} is specified, the crust-core phase transition 
can be inferred from the instability properties of the functional.
A first rough definition of the transition point
can be obtained from the spinodal instability which is associated to the liquid-gas phase transition in neutral nuclear matter at sub-saturation densities \cite{Ducoin07}. The spinodal region is defined as the convex part of the energy density in the neutron-proton density plane, and can be spotted from the presence of negative eigenvalues of the following curvature matrix: 
\begin{eqnarray}
C_{NM} 
 = \begin{pmatrix}
 \partial \mu_n/  \partial \rho_n &   \partial \mu_n/ \partial \rho_p \\
 \partial \mu_p/ \partial \rho_n &  \partial \mu_p/ \partial \rho_p
 \end{pmatrix}\label{eq:thspi}
\end{eqnarray} 
\\
i.e., the curvature matrix depends on the energy functional via the chemical potentials $\mu_q=\partial \epsilon/\partial n_q$, $(q=n,p)$.

The thermodynamic spinodal only gives a qualitative estimation of the transition point.
Indeed, in the presence of an electron background gas, the liquid-gas phase transition of 
nuclear matter is replaced by a transition from an homogeneous medium to a clusterized one.
Such a transition is better signalled by the instability of nuclear matter 
against independent fluctuations of neutron, proton and electron densities with a finite spatial 
extension\cite{Ducoin07,Pethick}. The energy density curvature matrix subject 
to finite size density fluctuations, takes the form

\begin{equation}
C_f =
\begin{pmatrix}
 \partial \mu_n/  \partial \rho_n &   \partial \mu_n/ \partial \rho_p & 0 \\
 \partial \mu_p/ \partial \rho_n &  \partial \mu_p/ \partial \rho_p & 0 \\
 0 &  0  &  \partial \mu_e/ \partial \rho_e \nonumber\\
 \end{pmatrix} 
 + k^2 \begin{pmatrix}
  C_{fin} & D_{fin} & 0\\
  C_{fin} & C_{fin} & 0\\
 0 & 0 & 0\\
 \end{pmatrix}
 + \frac{4 \pi e^2}{k^2} \begin{pmatrix}
 0 & 0 & 0\\
 0 & 1 & -1 \\
 0 & -1 & 1
 \end{pmatrix}\label{eq:dynspi}
 \end{equation}

In this matrix, in addition to the bulk term that corresponds to the thermodynamic spinodal, there is a surface term proportional to the 
square of the wavelength k, while the Coulomb interaction  adds a term 
inversely proportional to $k^2$.  Then thermodynamic fluctuations are recovered at 
$k \to \infty$ limit of this model. 
The global instability region is defined as the envelope of 
the different $k$-spinodals corresponding to all the possible linear sizes for the density fluctuation.

For both thermodynamic eq.(\ref{eq:thspi}) and dynamic eq.(\ref{eq:dynspi}) spinodals,
the crust-core transition point is defined as the intersection of the instability envelope
with the neutrino-less  $\beta$-equilibrium curve of 
neutron star matter, defined by:

\begin{equation}
    \mu_n-\mu_p=\mu_e.
\end{equation}
\\

\subsection{Introduction of the magnetic field}
\label{sec:mageos}
\indent It is well known that in the presence of strong magnetic fields, the motion of charged particles is confined to Landau levels, in a direction perpendicular 
to the magnetic field direction. This leads to a modification of the energy levels of the particles, and hence of the EoS of neutron star matter \cite{Chakrabarty}. 

In the formulation of the meta-modelling eq.(\ref{eq:e_hnm}), a modification of the single-particle levels directly affects the first term of the meta-functional, which physically corresponds to the kinetic energy per baryon in the non-relativistic approximation.
Within this approximation, the kinetic energy per baryon of neutrons remains unaltered, i.e.
\begin{equation}
e_{kin}^n = \frac{3 \hbar^2}{10 m_n^*} \left( \frac{3 \pi^2 n}{2} \right)^{2/3} (1 + \delta)^{5/3} \label{eq:ekinB}
\end{equation}
where $m_n^*$ from eq.(\ref{eq:effm_expn}) is the effective neutron mass. 
Conversely, electrons and protons are affected by the magnetic field. 
For the electrons, we follow the standard relativistic treatment of ref.\cite{eosBrel}.
Within the non-relativistic approximation \cite{eosBnonrel},  the single particle energy
of the $\nu$-th Landau level is given by
\begin{equation}
\mu_p^* = \frac{p_{z,F}^2}{2m_p^*} + \frac{\nu e B}{m_p^*}~,
\end{equation}
where $p_z$ is the momentum of the proton in the $z$-direction in which the magnetic field is directed. 
The proton density becomes modified in presence of the magnetic field as:
\begin{equation}
n_p = \frac{eB}{2 \pi^2} \sum_{\nu}^{\nu_{max}} g_{\nu} p_{z,F}(\nu)
\end{equation}
where the maximum number of Landau levels is given by: 
\begin{equation}
\nu_{max} = \frac{m_p^* \mu_p^*}{eB}.    
\end{equation} 
Then the proton contribution to the kinetic energy per baryon becomes:
\begin{equation}
e_{kin}^p = \frac{eB}{2n \pi^2} \frac{1}{m_p^*} \sum_{\nu}^{\nu_{max}} \left[ \frac{p_{z,F}^3(\nu)}{3} + 2\nu eB p_{z,F}(\nu) \right]~.
\end{equation}

When the magnetic field is high, only the lowest or a few Landau levels are occupied. The corresponding critical quantizing field $B_c$ can be calculated by equating the rest energy $mc^2$ of a particle of mass $m$ to its cyclotron energy $\hbar \omega_c$, where $\omega_c = eB/mc$ is the cyclotron frequency, i.e. $B_c = m^2 c^3/e \hbar$. For the case of electrons, this gives $B^e_c = 4.4 \times 10^{13}$ G. For low magnetic fields, the Landau levels are numerous and closely spaced, such that one recovers the continuum limit in the absence of the field.


\newpage
\section{Results: Magnetic field effects}
\label{sec:mag}

\subsection{Effect on the CCPT}
\label{sec:resspi}

We first analyze the effect of the magnetic field on the thermodynamical spinodal for our reference model SLy5. The results of this section are in   qualitative agreement with the previous works \cite{cp1,cp2,cp3} using relativistic mean-field functionals, and the Vlasov method to compute the dynamical spinodal.
\\

\begin{figure}[htbp]
  \begin{center}
      \includegraphics[width=.4\textwidth, angle=270]{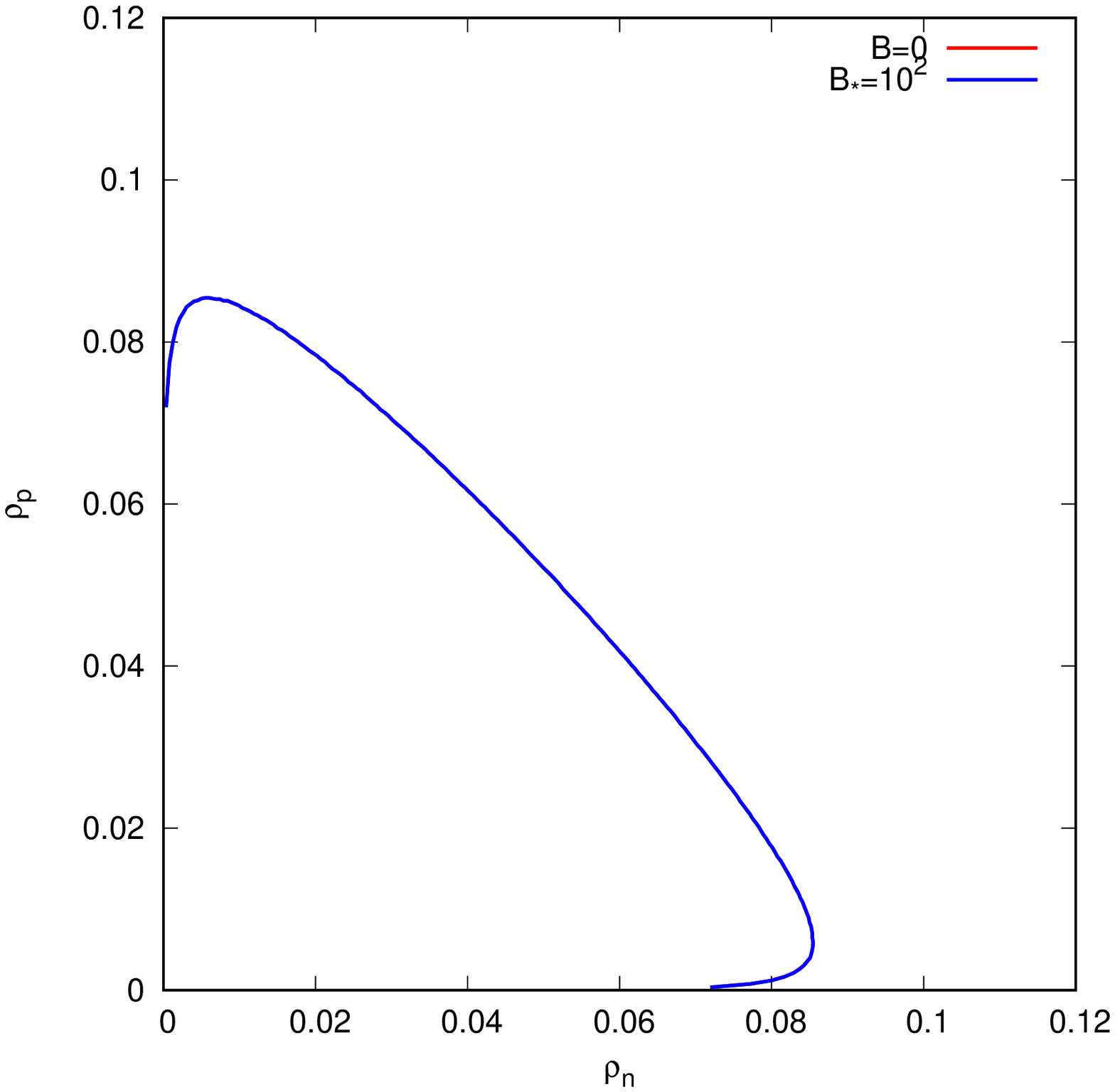}
      \includegraphics[width=.4\textwidth, angle=270]{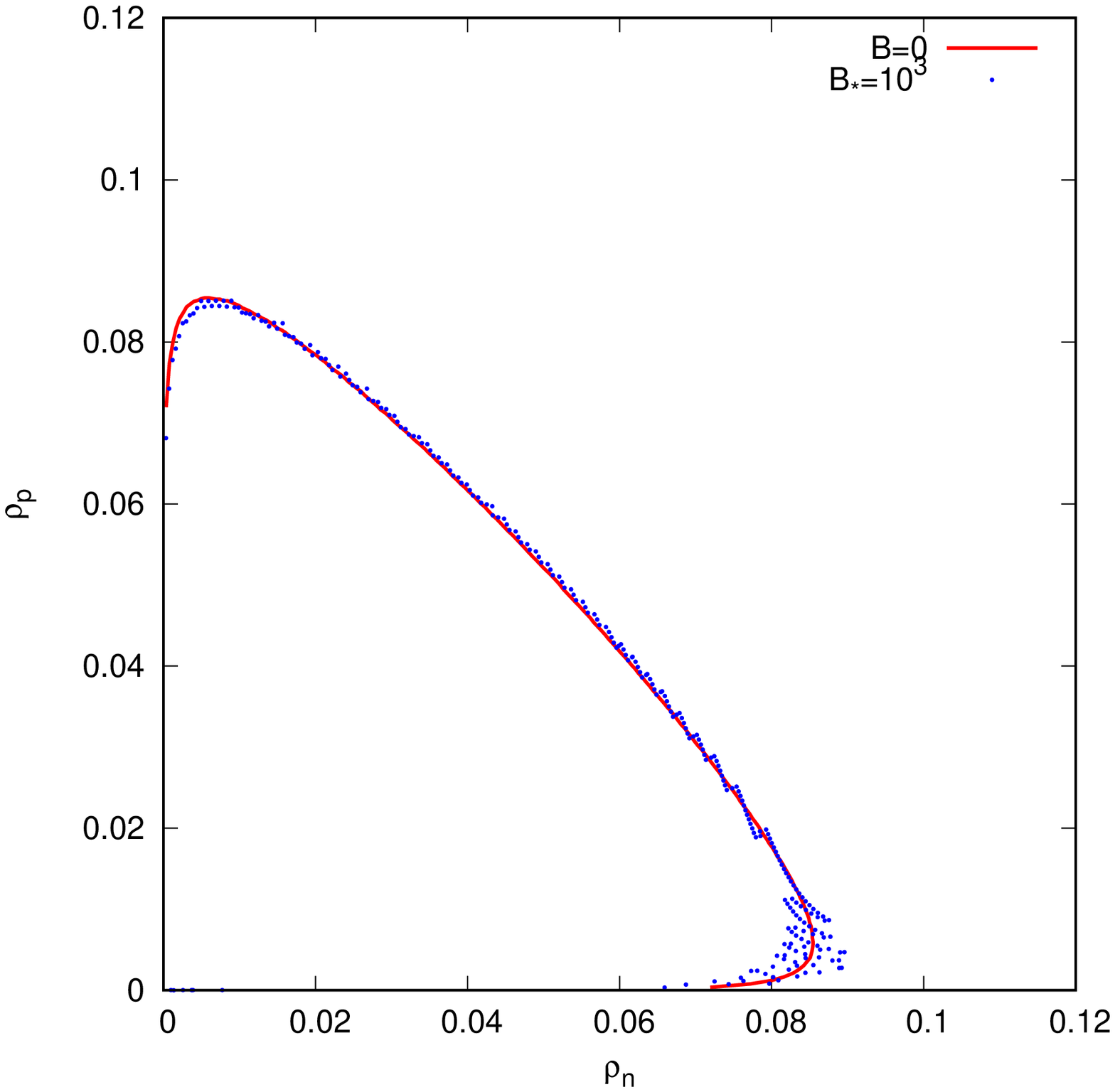}
      \includegraphics[width=.4\textwidth, angle=270]{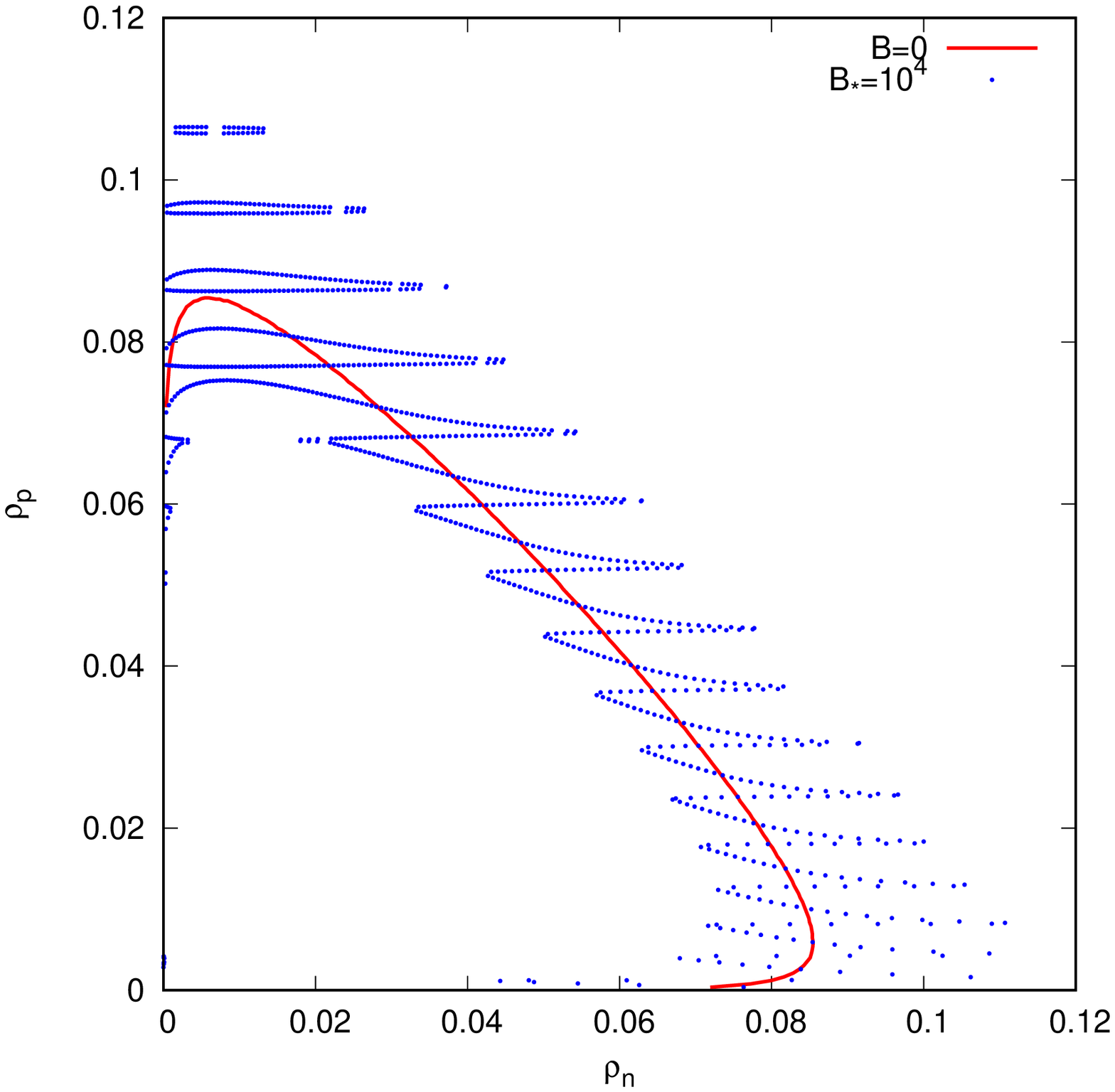}
      \includegraphics[width=.4\textwidth, angle=270]{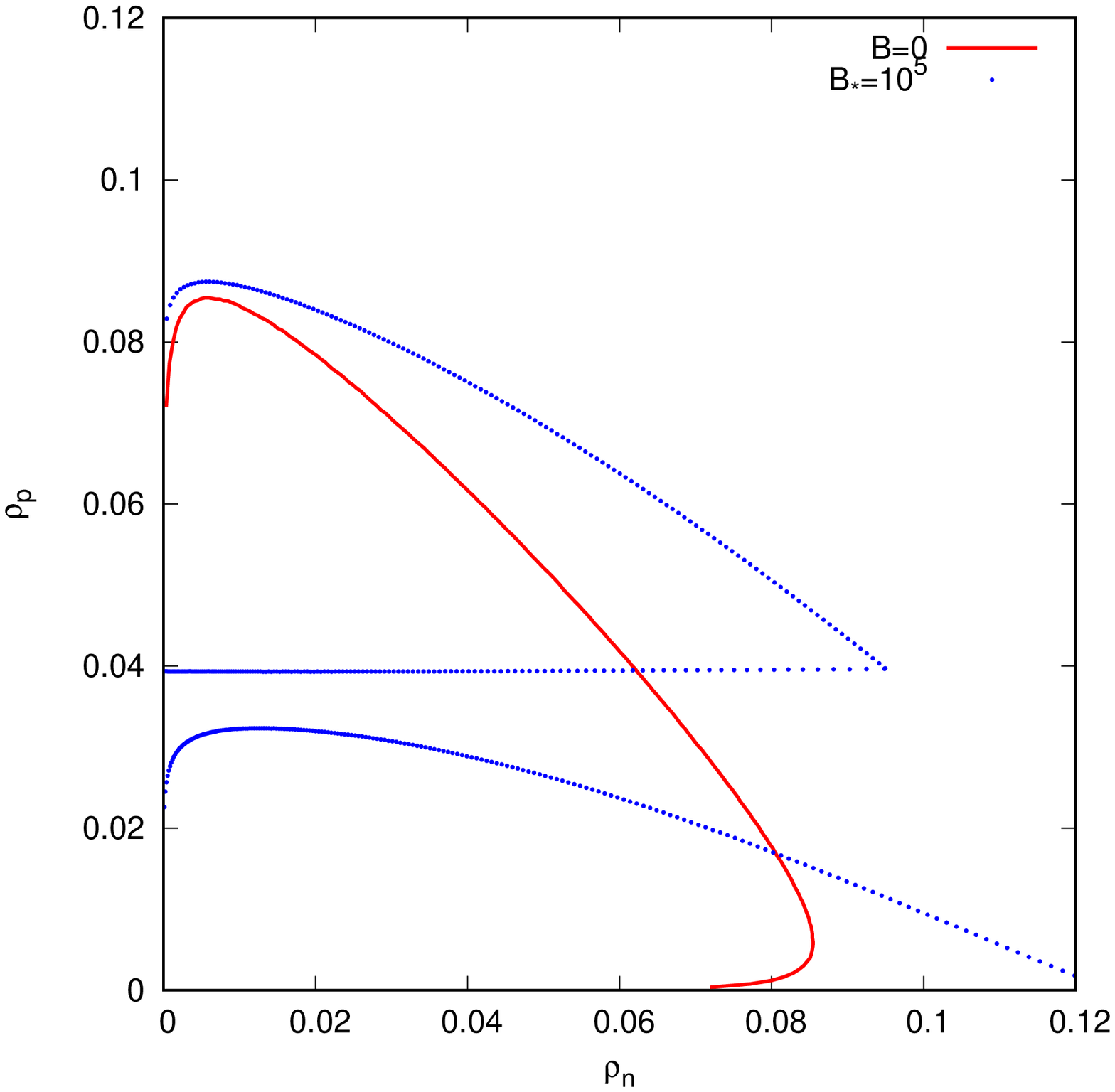}
    \caption{Thermodynamic spinodals for relative magnetic field strengths $B_* = 10^2$ 
    (upper left), $10^3$ (upper right), $10^4$ (lower left) and $10^5$ (lower right).The continuous red line gives the spinodal envelope in the absence of magnetic field.}
    \label{fig:tdspinbst1e2}
  \end{center}
\end{figure}

Taking multiples of the quantizing electron field $B^e_c = 4.4 \times 10^{13}$ G, we express the magnetic field in units of $B_* = B/B^e_c$.
The effect of the magnetic field on the thermodynamical spinodal is shown in Fig.
\ref{fig:tdspinbst1e2}. It is found to be negligible up to $10^2 B^e_c$. However, for fields above $10^3 B^e_c$,
the effect of magnetic field on the thermodynamical spinodal region is quite dramatic.  Instead of a single region, we now find alternate regions of stability and instability. 
This is due to the appearance of kinks in the EoS as a result of the confinement of the motion of the charged particles to quantized Landau levels. 
\\

We now turn to investigate how magnetic field affects the CCPT for the thermodynamical and dynamical spinodals.  As already discussed in section \ref{sec:ccpt}, the CCPT is determined from the crossing of the spinodals with the 
$\beta$-equilibrium line.  \\

\begin{figure}[htbp]
  \begin{center}
      \includegraphics[width=.45\textwidth, angle=270]{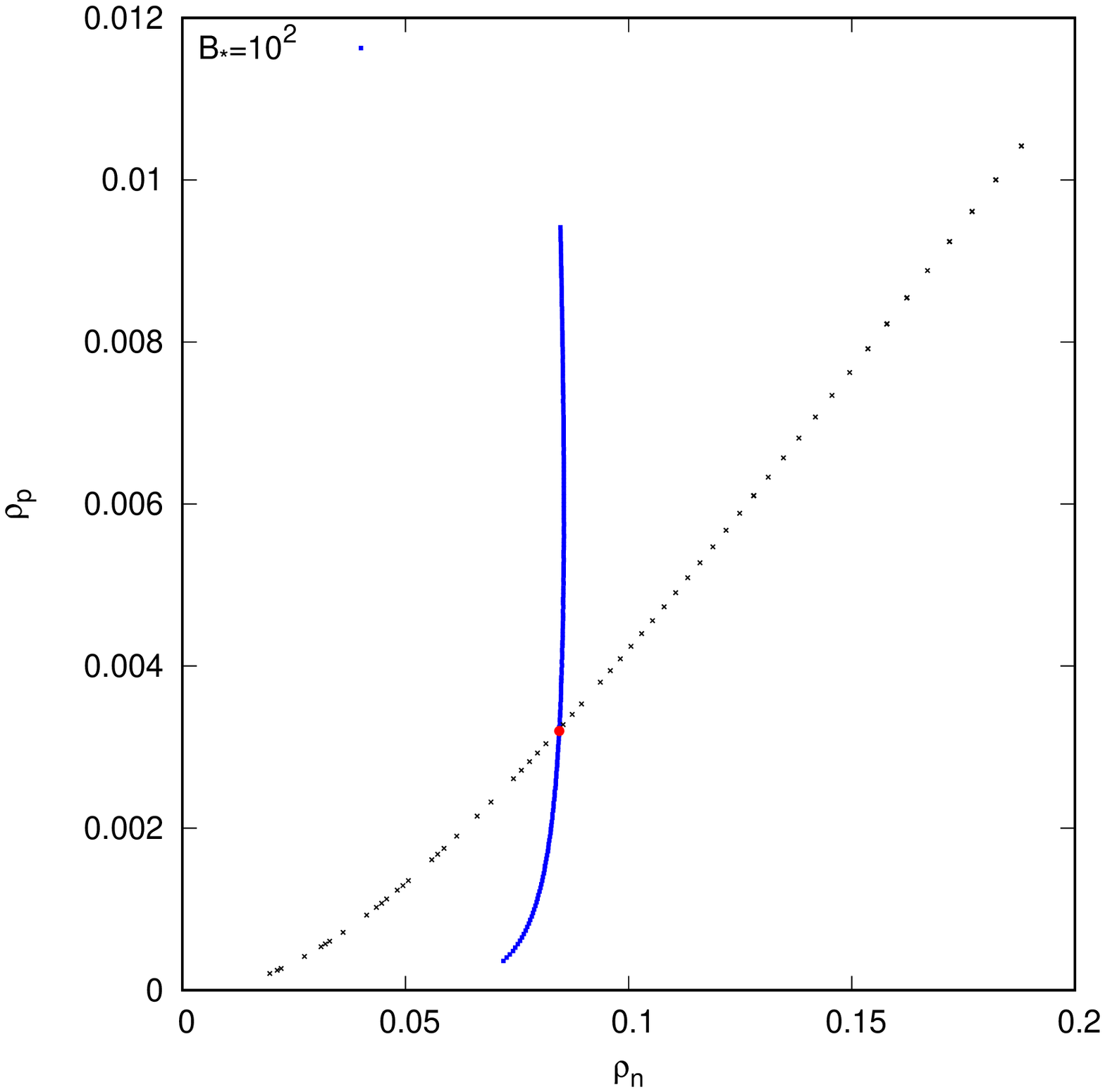}
      \includegraphics[width=.45\textwidth, angle=270]{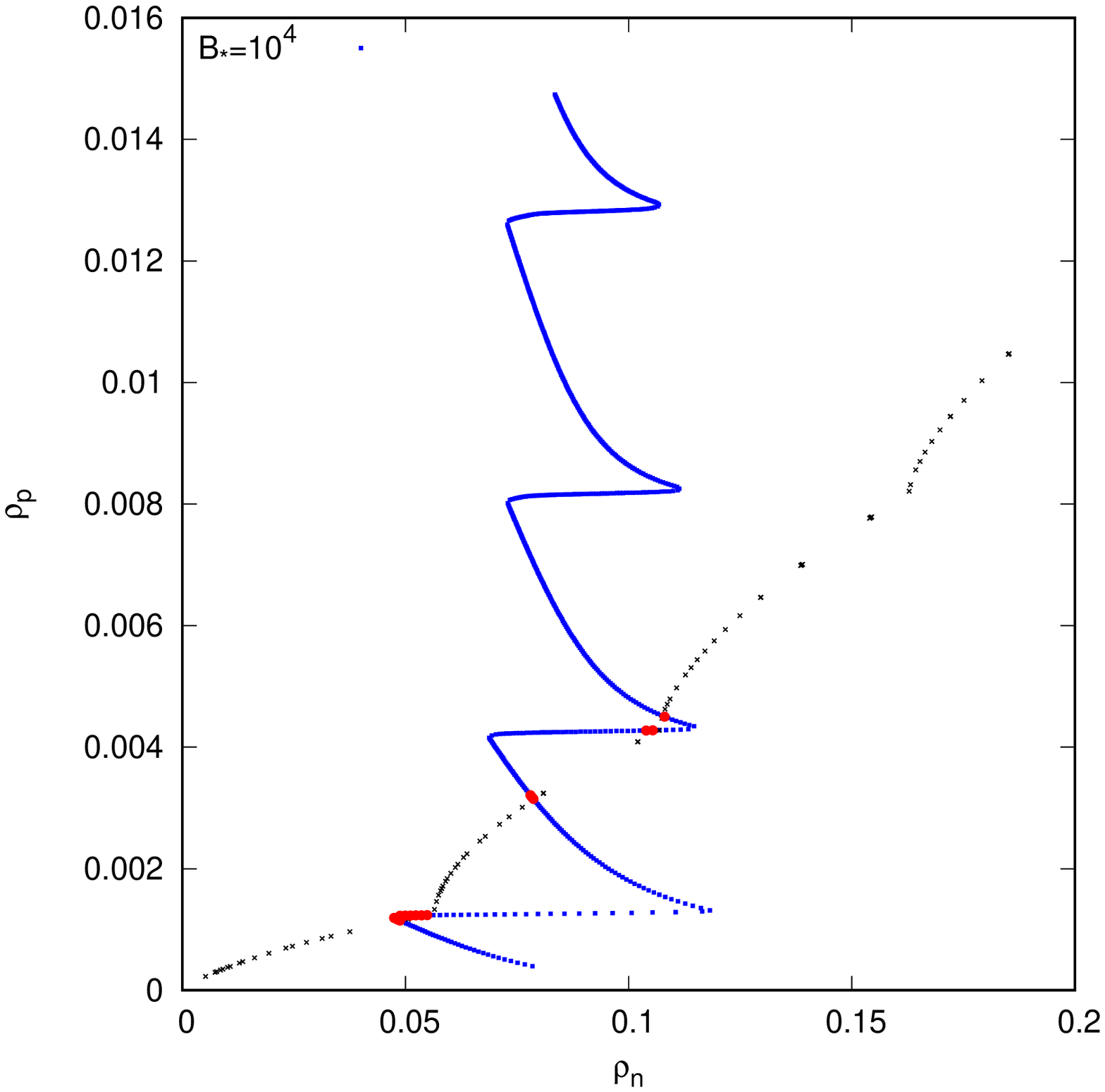}
      \includegraphics[width=.45\textwidth, angle=270]{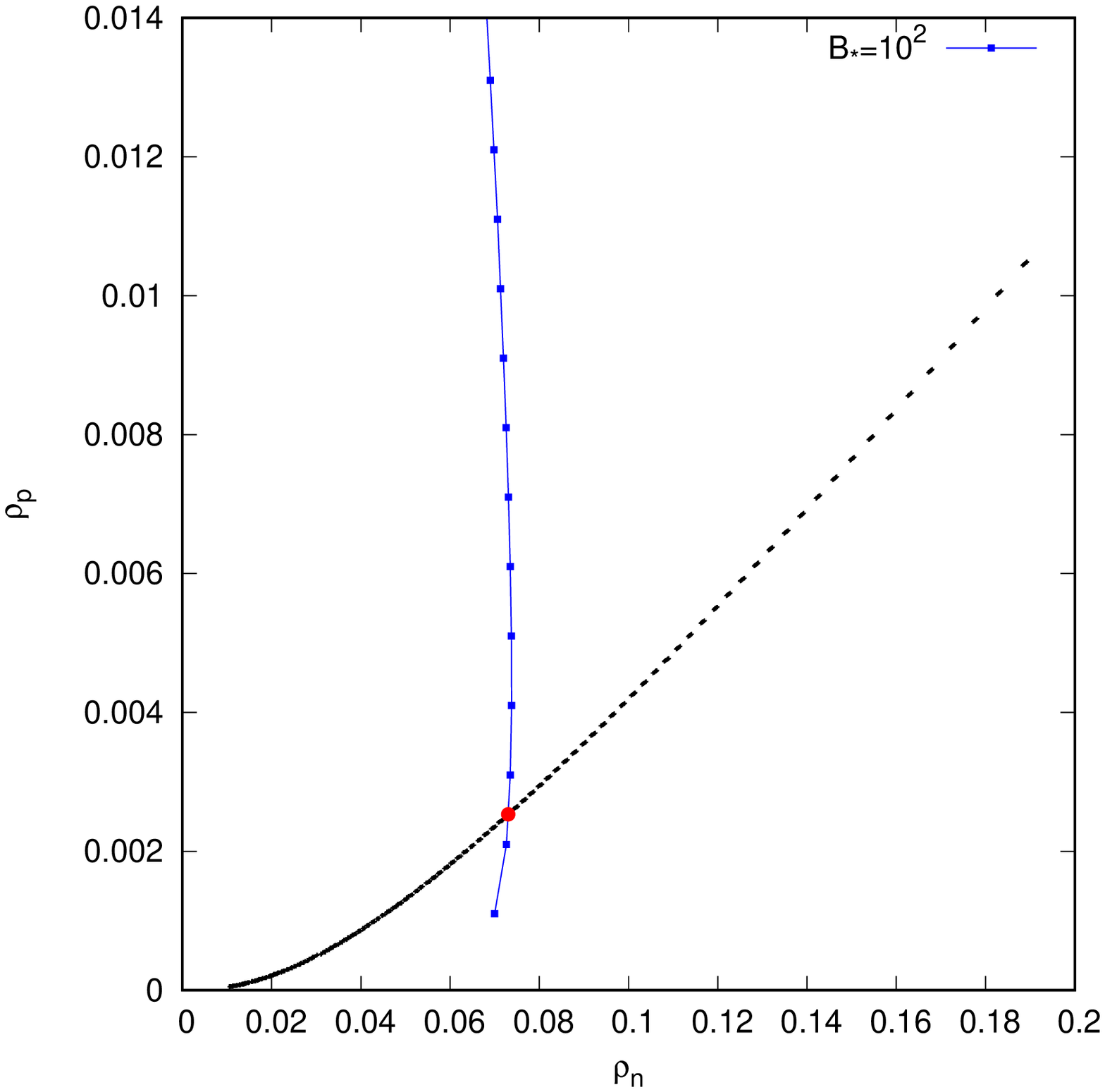}
      \includegraphics[width=.45\textwidth, angle=270]{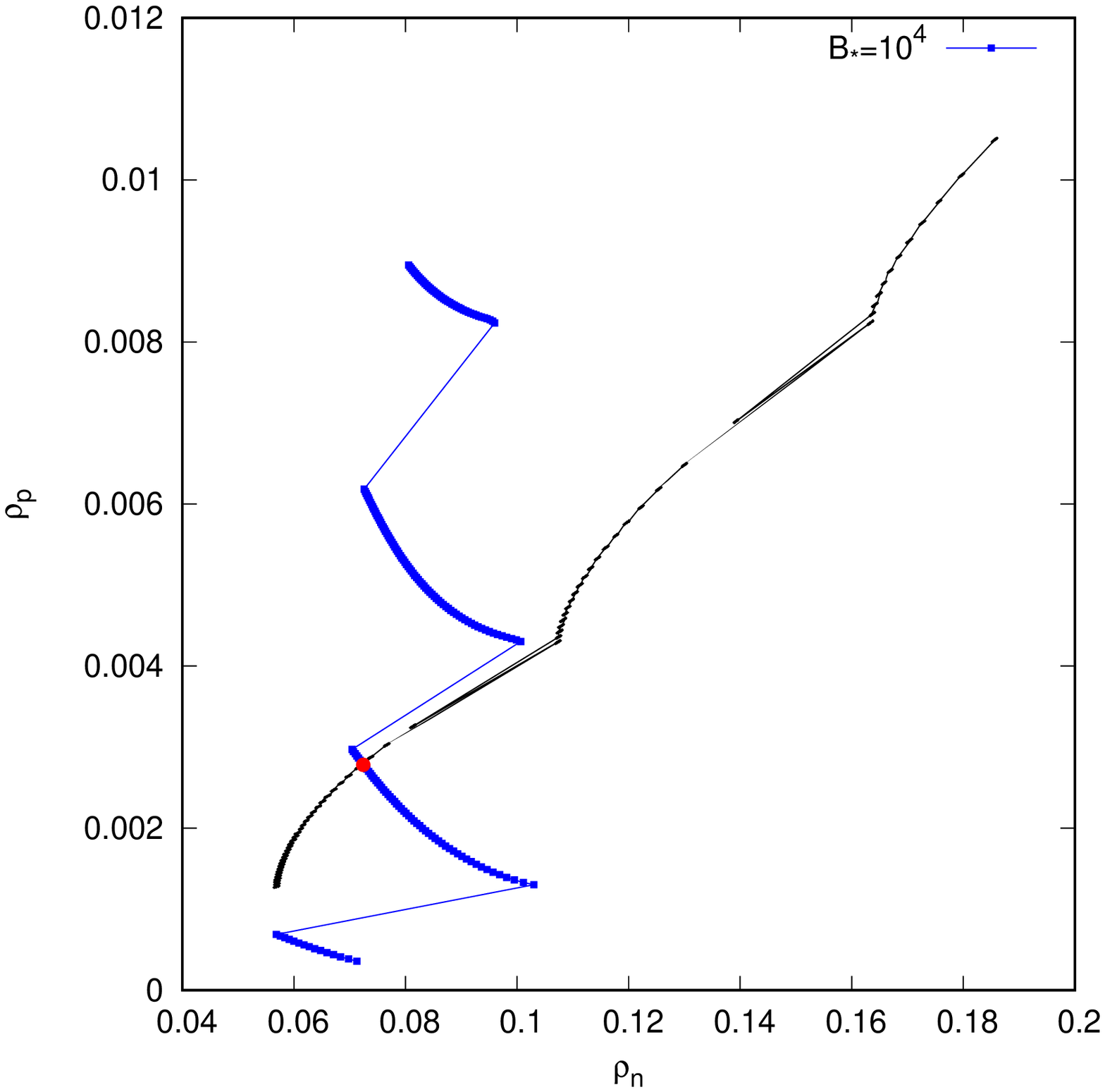}
    \caption{The crossing of the $\beta$-equilibrium line with the thermodynamic (upper panels) and dynamical (lower panels) spinodal for relative magnetic field strength $B_* = 10^2$ (left column) and $10^4$ (right column). The SLy5 functional is used.}
    \label{fig:beta_eossly5_bst1e2}
  \end{center}
\end{figure}

In Figure
\ref{fig:beta_eossly5_bst1e2}
the solid blue lines determine the spinodals,
the $\beta$-equilibrium lines are shown by the black points, while the crossing points are demarcated by the red dots. We 
can see that the qualitative behavior of the two spinodals is very similar, but the transition occurs at lower density when finite size fluctuations are taken into account (lower panels), as expected.
Because of the non-monotonous behavior of the spinodals in the presence of magnetic fields, 
it is clear that several crossing points are possible. This is however not a general rule, as we can 
see from the lower right part of Fig. \ref{fig:beta_eossly5_bst1e2}: a single transition point is observed even at very high magnetic fields for the dynamical spinodal of SLy5. This is at variance with the findings of ref.\cite{cp1,cp2,cp3} and shows that the possibility of having successive layers of stable and unstable matter is model dependent.
\\

In Tables \ref{tab:lgpt_mag} and \ref{tab:ptdyn_mag}, we list the first and the last 
crossing points as obtained from the thermodynamic and the dynamic analysis,
as a function of the magnetic field. This defines the density $\Delta \rho_{PT}$ and pressure $\Delta p_{PT}$ 
interval corresponding to the successive sequence of stable and unstable thermodynamic configurations, due to the presence of Landau levels.

\begin{table*}[htbp]
   \caption{The uncertainty in the transition point for thermodynamical spinodals of the SLy5 functional, for different magnetic field strengths (see text for more details).}
\begin{tabular}{|c|c|c|c|c|c|c|}
\hline
   $B^*$ & $\rho_1$ & $\rho_2$ & $\Delta \rho_{PT}$ & $p_1$ & $p_2$ & $\Delta p_{PT}$\\
    {} & ($fm^{-3}$) & ($fm^{-3}$) & ($fm^{-3}$) & ($MeV/fm^{-3}$) & ($MeV/fm^{-3}$) & ($MeV/fm^{-3}$)\\
 \hline
 \hline
 $10^2$ & 0.088 & 0.088 & 0. & 0.462 & 0.462 & 0. \\
 \hline
 $10^3$ & 0.086 & 0.091 & 0.005 & 0.442 & 0.502 & 0.06\\
\hline
$5 \times 10^3$ & 0.075 & 0.112 & 0.037 & 0.332 & 0.882 & 0.550\\
\hline
$7 \times 10^3$ & 0.070 & 0.090 & 0.020 & 0.300 & 0.501 & 0.201\\
\hline
$10^4$ & 0.048 & 0.113 & 0.065 & 0.193 & 0.937 & 0.744 \\
\hline
\end{tabular}
\label{tab:lgpt_mag}
\end{table*}

As previously observed in ref.\cite{cp1,cp2,cp3},  the global effect of the magnetic field is to increase the density and pressure region where the EoS behavior is not monotonic, but the effect is non-linear. We observe that this effect is somewhat quenched in the dynamical spinodal treatment with respect with the thermodynamical one.

\begin{table*}[htbp]
   \caption{Density and pressure of the transition point for dynamical spinodals of the SLy5 functional, for different magnetic field strengths (see text for more details).}
\begin{tabular}{|c|c|c|c|c|c|c|}
\hline
   $B^*$ & $\rho_1$ & $\rho_2$ & $\Delta \rho_{PT}$ & $p_1$ & $p_2$ & $\Delta p_{PT}$\\
    {} & ($fm^{-3}$) & ($fm^{-3}$) & ($fm^{-3}$) & ($MeV/fm^{-3}$) & ($MeV/fm^{-3}$) & ($MeV/fm^{-3}$)\\
 \hline 
 \hline
 $10^2$ & 0.079 & 0.079 & 0. & 0.356 & 0.356 & 0. \\
 \hline
 $10^3$ & 0.075 & 0.077 & 0.002 & 0.320 & 0.338 & 0.018\\
\hline
$5 \times 10^3$ & 0.071 & 0.085 & 0.014 & 0.297 & 0.438 & 0.141\\
\hline
$7 \times 10^3$ & 0.065 & 0.085 & 0.020 & 0.261 & 0.439 & 0.178\\
\hline
$10^4$ & 0.043 & 0.077 & 0.034 & 0.165 & 0.369 & 0.204 \\
\hline
\end{tabular}
\label{tab:ptdyn_mag}
\end{table*}

\subsection{Sensitivity of CCPT to influential parameters}
\label{sec:sensrho_dyn}

It was shown by different authors \cite{Ducoin2008,Ducoin2011,Xu2009,Pearson2012} that the core-crust phase transition in ordinary non-magnetized neutron stars strongly depends on the symmetry energy, particularly on the slope parameter $L_{sym}$.  Within the MM approach, in a recent analysis \cite{Antic} a systematic study was performed in order to isolate the effects of the empirical nuclear parameters and their uncertainties. It was found that the most influential parameters are the poorly known isovector parameters $L_{sym}$ and $K_{sym}$.  
In this section, we study the effect of  $L_{sym}$ and $K_{sym}$ on the CCPT density as estimated from the thermodynamic spinodal, as a function of the magnetic field. \\

We take as fiducial values the ones of the SLy5 functional, $L_{sym}$ = 48.3 MeV and $K_{sym}$ = -112 MeV.
To study the sensitivity to those parameters, we modify the parameters one by one keeping all the others unchanged. For the parameter under study, we consider two extreme values corresponding to the present uncertainty from nuclear physics experiments as estimated in ref. \cite{Casali1}, namely $L_{sym}^{min}$ = 47 MeV, $L_{sym}^{max}$ = 106 MeV,  $K_{sym}^{min}$ = -135 MeV,  $K_{sym}^{max}$ = 129 MeV.
\\

\begin{figure}[htbp]
  \begin{center}
      \includegraphics[width=.45\textwidth, angle=270]{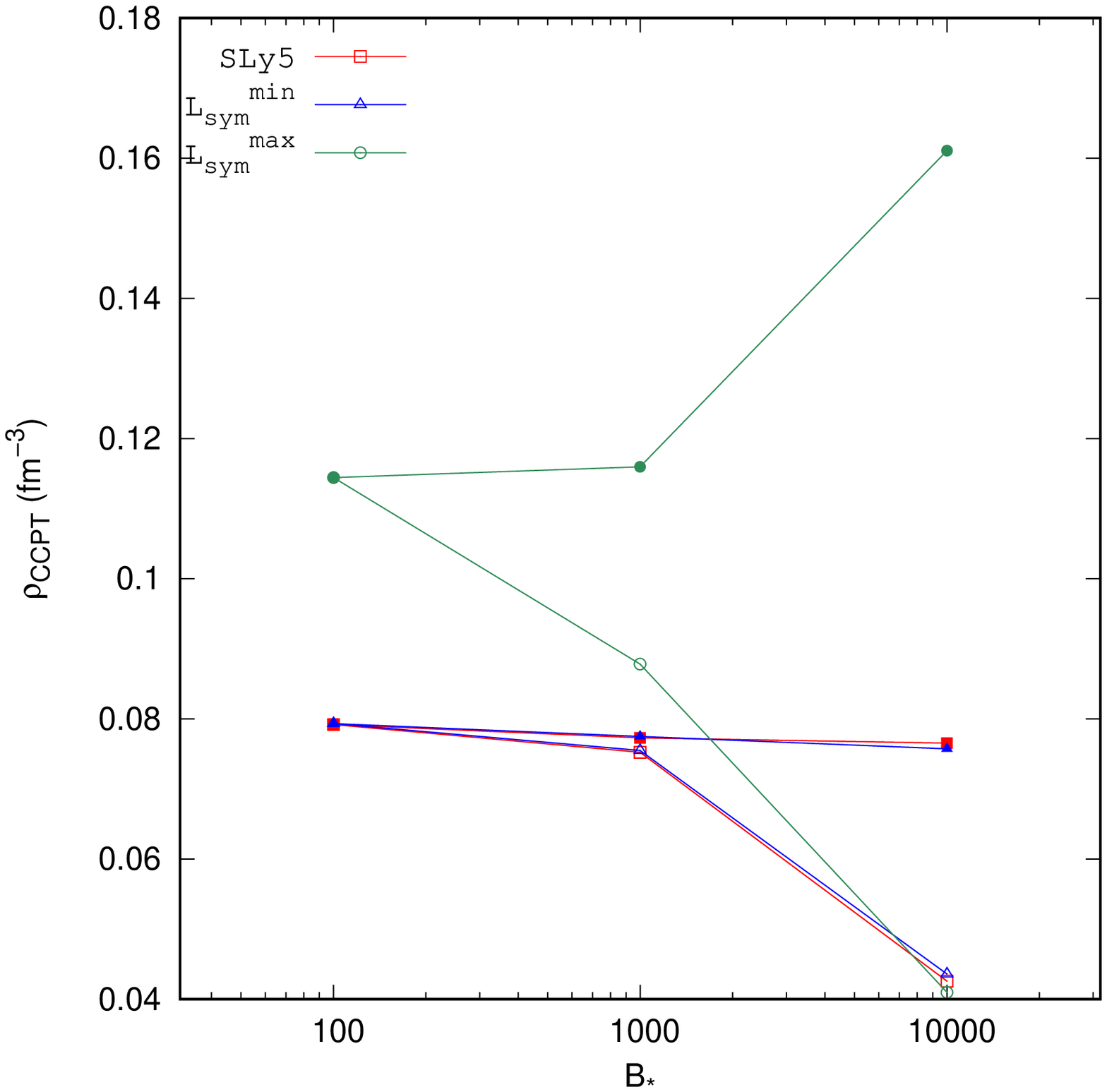}
      \includegraphics[width=.45\textwidth, angle=270]{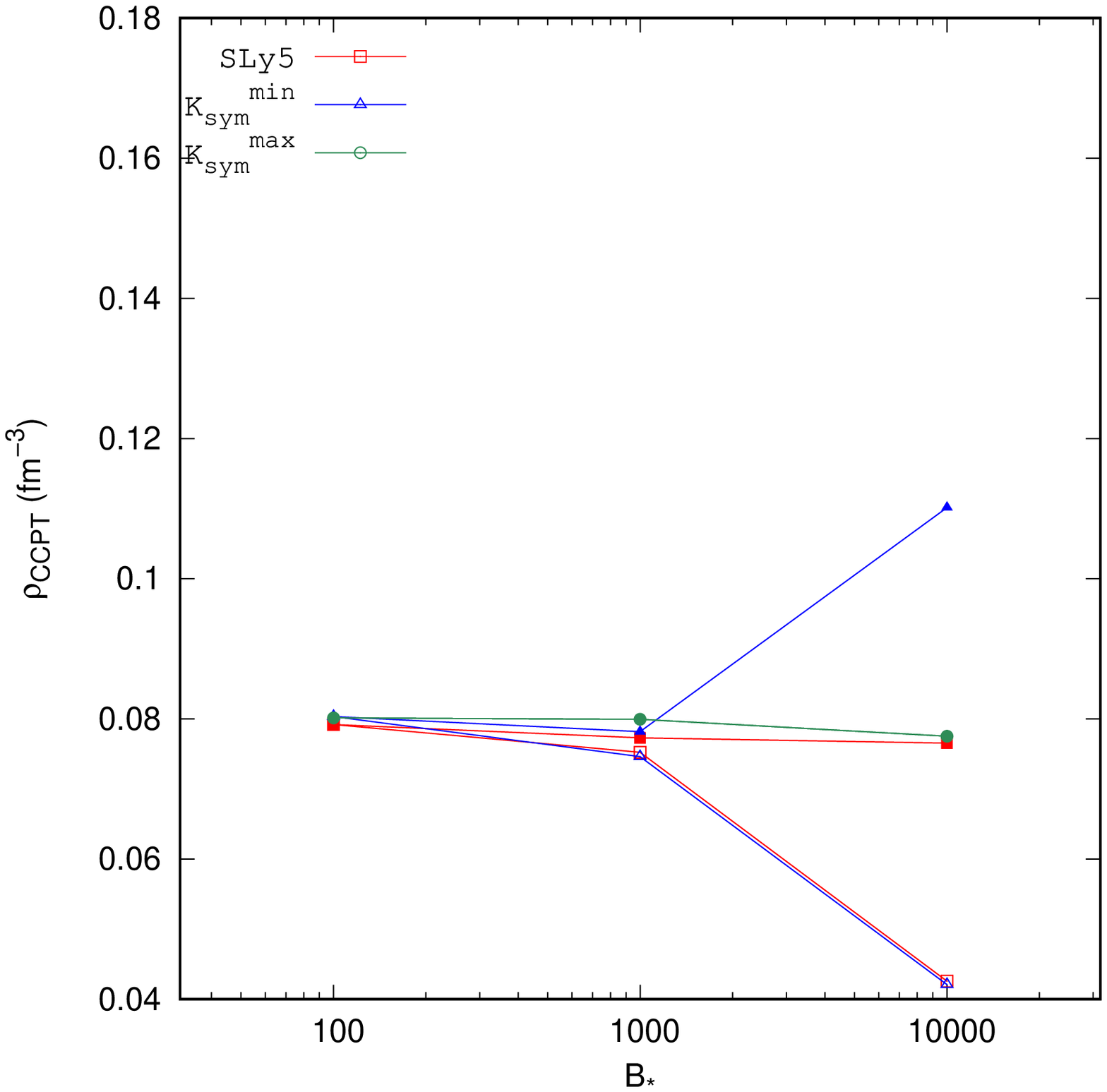}
   \caption{Sensitivity of the CCPT density from the dynamical spinodal analysis to $L_{sym}$ (left) and $K_{sym}$ (right). Open (filled) symbols represent the first (last) crossing of the beta-equilibrium line with the spinodal envelope.}
    \label{fig:sensrho_lsym2}
  \end{center}
\end{figure}

\begin{figure}[htbp]
  \begin{center}
      \includegraphics[width=.45\textwidth, angle=270]{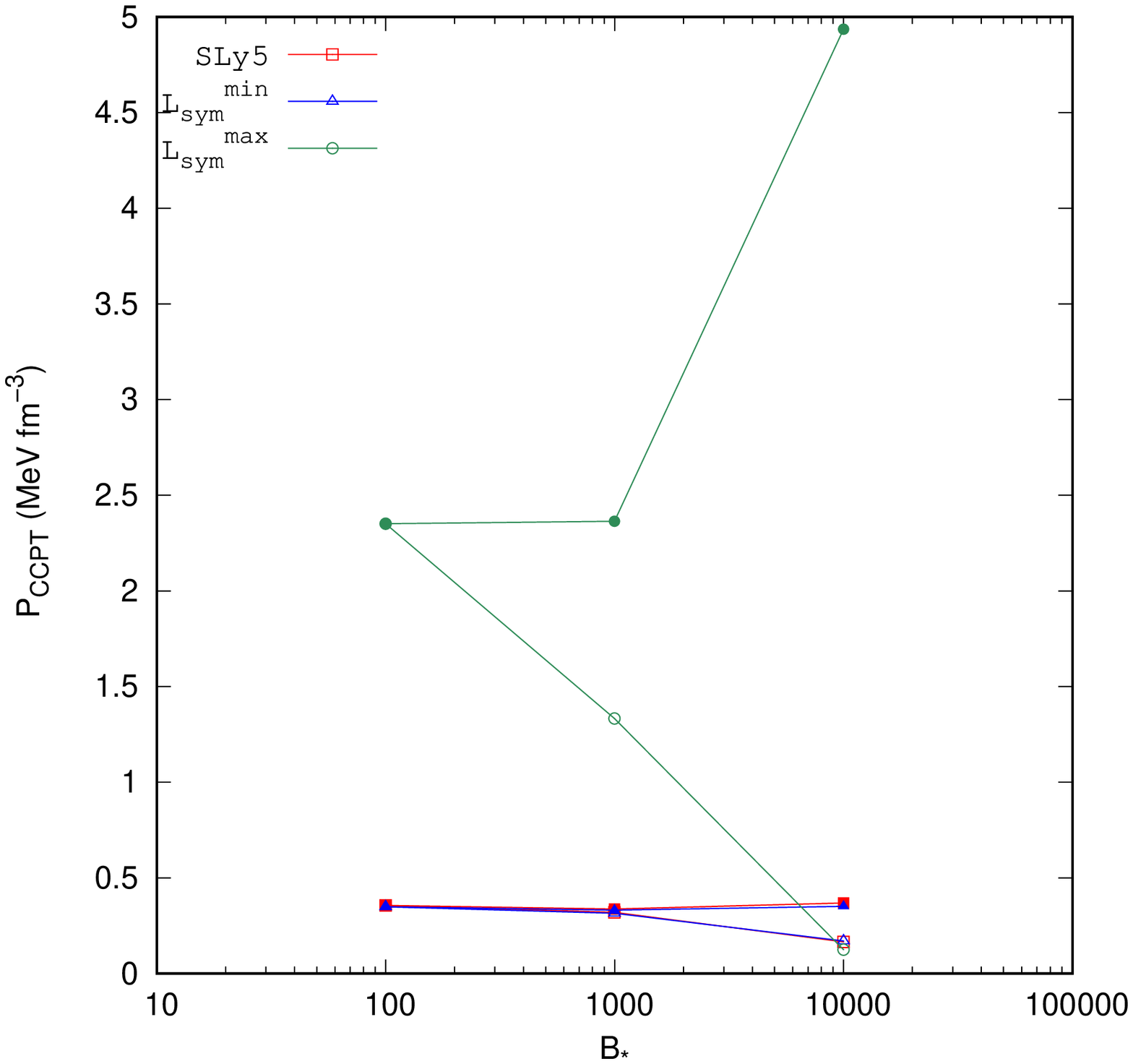}
      \includegraphics[width=.45\textwidth, angle=270]{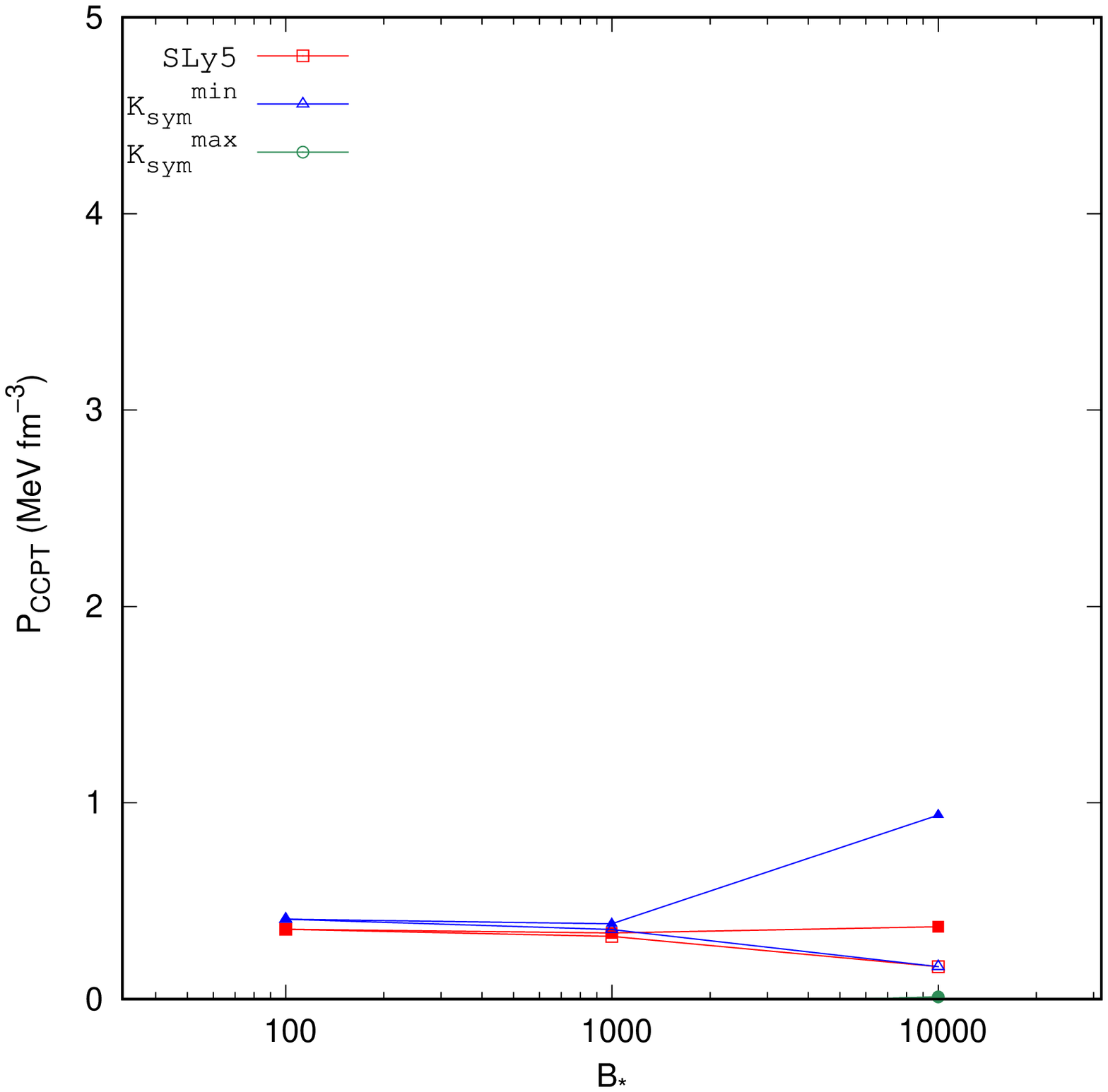}
   \caption{Sensitivity of the CCPT pressure from the dynamical spinodal analysis to $L_{sym}$ (left) and $K_{sym}$ (right). Open (filled) symbols represent the first (last) crossing of the beta-equilibrium line with the spinodal envelope.}
    \label{fig:senspres_lsym2}
  \end{center}
\end{figure}
 
We can see from the results of Figures \ref{fig:sensrho_lsym2} and \ref{fig:senspres_lsym2} that the transition point is strongly modified by a change of the slope parameter $L_{sym}$. As a general statement, a stiffer symmetry energy leads to higher values for the transition density, as already observed in the literature \cite{Ducoin2008,Ducoin2011,Xu2009,Pearson2012}. We can additionally observe that the effect strongly increases for the highest magnetic fields.
For small values of $L_{sym}$ such as SLy5 or $L_{sym}^{min}$, the transition densities $\rho_1$ and $\rho_2$ coincide, 
but for large values of $L_{sym}$ they are quite different for different magnetic field strengths $B_*$ .
In particular, we can see that the existence of successive stable and instable regions around the transition point seems to be strongly correlated with the $L_{sym}$ parameter.
Concerning  $K_{sym}$, we can see that it is by far less influential than $L_{sym}$. Variations of $K_{sym}$ in the physically reasonable range $K_{sym}^{min}$ to $K_{sym}^{max}$ 
does not change the CCPT density considerably for different magnetic field strengths $B_*$. An important effect is only seen for the highest magnetic field, when two different transition points start to appear due to the highly complex behavior of the spinodal zone. 
\\

From Figure \ref{fig:sensrho_lsym2} we can conclude that better constraints on the symmetry energy slope $L_{sym}$ are expected to strongly reduce the model dependence of the results.
\\

\section{Magnetic field effects on the mass-radius relation}
\label{sec:massradius}

In the previous section, we demonstrated how much a strong magnetic field affects the thermodynamical and dynamical spinodals
and hence the crust-core phase transition in neutron stars.  
We can therefore expect following the previous studies \cite{cp1,cp2,cp3} that the magnetic field will also play
an important role in determining the structure of neutron stars, especially its radius and crust thickness. In order to investigate the influence
of magnetic field on the static properties of strongly magnetized neutron stars,
we choose as a reference model for this section the Skyrme SLy4 EoS, 
which is one of the most commonly used EoS in astrophysics  \cite{DouchinHaensel}. 
\\

Structure calculations of strongly magnetized neutron stars have been discussed intensively in the literature in the past few years
\cite{LopesMenezes, MenezesLopes, Dexheimer2011, MenezesAlloy, Dexheimer2017, Chatterjee2014,Chatterjee2016}. It is well known that strong electromagnetic field affects neutron stars in two ways:
firstly the EoS due to the quantization of the charged particles into Landau levels, and secondly by introducing an anisotropy in the 
energy momentum tensor and thereby affecting the global structure.
In the presence of strong magnetic fields, the neutron star shape strongly deviates from spherical symmetry and hence the spherically symmetric Tolman-Oppenheimer-Volkov
(TOV) equations are no longer applicable. The ideal method to tackle the problem  is to self-consistently solve the neutron star structure equations endowed with a magnetic field, i.e. 
Einstein--Maxwell and equilibrium equations, with a magnetic field dependent EoS. 
However this is a complicated numerical problem that has only been tackled recently \cite{Chatterjee2014,Chatterjee2016}.
In this section, we compute  and compare the mass-radius relations of magnetized neutron stars obtained  using both an isotropic TOV solution \cite{cp1,cp2,cp3}, as well as a full numerical solution.

\subsection{TOV calculation}
\label{sec:tov}

In the calculation of the core EoS, the magnetic field contribution to the energy density and pressure,
$\epsilon_B = p_B = B^2/8 \pi$, is assumed to be isotropic :
\begin{eqnarray}
\epsilon_{tot} &=& n e_{HNM} + \epsilon_l + \epsilon_B \, ; \nonumber \\
p_{tot} &=& p_{HNM} + p_l + p_B \; .
\end{eqnarray}
Here, $e_{HNM}$ ($p_{HNM}$) is the hadronic contribution to the energy (pressure) from the MM EoS eq.(\ref{eq:e_hnm}), modified such as to include the magnetic field contribution as in eq.(\ref{eq:ekinB}). 
The lepton contribution $\epsilon_l$, $p_l$ is calculated considering electrons as a relativistic free Fermi gas, again modified by the magnetic field as explained in section \ref{sec:mageos}.

Using the core EoS and the CCPT points calculated in the previous subsection for the SLy4 case, we aim to
compute the mass-radius relations of magnetized neutron stars. The calculation of the structure of neutron stars using TOV
requires the additional knowledge of the crust EoS. This can however be avoided by using the method proposed 
by Zdunik et al. \cite{Zdunik}. In this method, one integrates from the core to the crust-core interface following the TOV prescription 
in order to obtain the core mass and radius. A simple algebraic calculation relates  the crust mass $M_{crust}$ and the total radius $R_{NS}$ 
 with the core mass $M_{core}$ and core radius $R_{core}$ and the CCPT density $\rho_{cc}$ and pressure $P_{cc}$ as:
\begin{eqnarray}
M_{crust} &=& \frac{4 \pi P_{cc} R^4_{core}}{G M_{core}} \left( 1 - \frac{2 G M_{core}}{R_{core} c^2} \right) \nonumber \\
R_{NS} &=& \frac{R_{core}}{1-(\alpha-1)(R_{core} \> c^2 \> G \> M_{NS} - 1) } \nonumber 
\end{eqnarray}
where $\alpha = \left (\frac{\mu_{cc}}{\mu_0}\right)^2 $ is the ratio of the baryon chemical potentials at the crust-core interface and that of cold catalyzed
matter at zero pressure (930.4 MeV). The crust thickness is then easily obtained from the relation $l_{crust} = R_{NS} - R_{core}$. \\
This approximation was shown to be reasonably precise for not too low neutron star masses in ref. \cite{Zdunik}.

\begin{figure}[htbp]
  \begin{center}
      \includegraphics[width=.6\textwidth, angle=270]{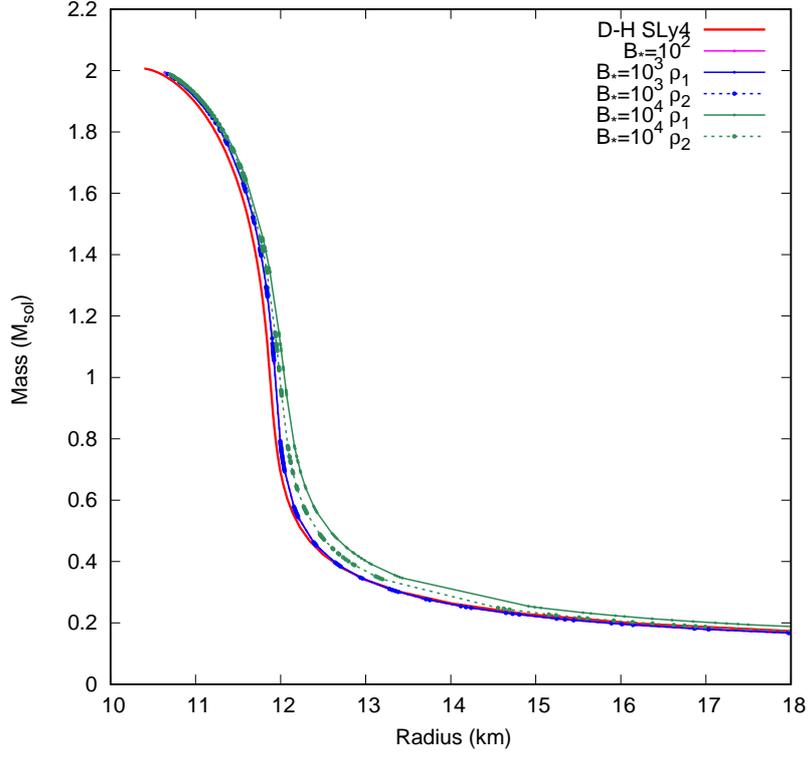}
    \caption{Mass-radius relation of neutron stars endowed with various magnetic field strengths computed using the isotropic TOV approximation and the Zdunik method for the SLy4 EoS.
    For comparison the mass-radius relation using the Douchin Haensel SLy4 EoS for B=0 from ref.\cite{DouchinHaensel} is also plotted.}
    \label{fig:mrcompzdunik_sly4_allb}
  \end{center}
\end{figure}

In Fig.~\ref{fig:mrcompzdunik_sly4_allb}
the mass-radius relation of neutron stars endowed with different magnetic field strengths computed within the spherical TOV approximation using the Zdunik method for the 
    SLy4 EoS are shown. Notice that the maximum mass is not always reached, a limitation of the meta-modelling due to the fact that the convergence of the Taylor expansion is not perfect at the highest densities. However, this is not a problem here, since we are mainly interested in the radius, which depends mainly on the low density region.
    For comparison the mass-radius relation using the Douchin Haensel (DH) SLy4 EoS \cite{DouchinHaensel} for B=0 is also plotted. This EoS is based on the same SLy4 energy functional, but the inhomogeneous crust contribution is explicitly accounted for. The close agreement between the DH result and our result at the lowest magnetic field demonstrated the quality of the Zdunik approximation. We can also see  from the 
    figure that except the case of very strong magnetic fields $B_* \sim 10^{4}$, the effect of magnetic field 
    on the total mass and radius of the neutron star is almost negligible in the spherical TOV approximation, in agreement with the results of ref.\cite{cp1,cp2,cp3}. However,the
    magnetic field has a non-negligible effect on the dynamical spinodals and hence the crust-core phase transition, which in turn
    should affect the crust-thickness. In Table~\ref{tab:table_lcrust}, we summarize the effect of different magnetic field strengths
    on the transition densities and the neutron star structure. For zero or low magnetic fields ($B_* \sim 10^{2}, 10^{3}$), there is a unique transition density.
    However for higher fields, as discussed in Sec.~\ref{sec:resspi}, the $\beta$-equilibrium line crosses the dynamical spinodals
    at multiple points. The transition density corresponding to the first and the last crossings are denoted as $\rho_1$ and 
    $\rho_2$ in the table. The difference in crust-core transition density leads to a difference in the corresponding total 
    and core radii and crust thickness $l_{crust} = R_{NS} - R_{core}$. The values of $l^1_{crust}$ and $l^2_{crust}$, 
    corresponding to $\rho_1$ and $\rho_2$, are also given in the table, along with their difference 
    $\Delta l_{crust} = l^2_{crust}- l^1_{crust}$. In this case, both $l^1_{crust}$ and $l^2_{crust}$ are of the order of 1 Km, much thinner than what was suggested in \cite{jorge2014}. $\Delta l_{crust}$ measures the region in the neutron star crust where layers of homogeneous and inhomogeneous matter might coexist \cite{cp1,cp2,cp3}. These results are in good qualitative agreement with ref.\cite{cp1,cp2,cp3}, even if some quantitative differences arise from the use of a different functional. The model dependence of the results will be discussed in section \ref{sec:models}.

\begin{table*}[htbp]
   \caption{Effect of a strong magnetic field on the transition density and crust thickness for a neutron star of 
   gravitational mass 1.4 M$_{\odot}$ with the SLy4 functional, in the isotropic TOV approximation. See text for details. }
\begin{tabular}{|c||c|c|c|c||c|c|c|c||c|}
\hline
 & & & & & & & & & \\
    $B_*$ & $\rho_1$ & $R^1_{NS}$ & $R^1_{core}$ & $l^1_{crust}$ & $\rho_2$ & $R^2_{NS}$ & $R^2_{core}$ & $l^2_{crust}$ &  $\Delta l_{crust}$  \\
    {} & (fm$^{-3}$) & (km) & (km) & (km) & (fm$^{-3}$) & (km) & (km) & (km) & (km)\\
    \hline 
 \hline
 $0$ & 0.077 & 11.704 & 10.797 & 0.908 & 0.077 & 11.704 & 10.797 & 0.908 & 0 \\
 \hline
 $10^2$ & 0.076 & 11.771 & 10.843 & 0.928 & 0.076 & 11.771 & 10.843 & 0.928 & 0 \\
 \hline
 $10^3$ & 0.071 & 11.770 & 10.859 & 0.911 & 0.074 & 11.770 & 10.849 & 0.920 & 0.009  \\
  \hline
 $5 \times 10^3$ & 0.070 & 11.764 & 10.865 & 0.899 & 0.084 & 11.761 & 10.8 & 0.961 & 0.062 \\
  \hline
 $7 \times 10^3$ & 0.033 & 11.804 & 10.993 & 0.811 & 0.081 & 11.777 & 10.811 & 0.966 & 0.155 \\
  \hline
 $10^4$ & 0.041 & 11.826 & 10.97 & 0.856 & 0.074 & 11.795 & 10.855 & 0.940 & 0.084 \\
\hline
\end{tabular}
\label{tab:table_lcrust}
\end{table*}

One may also calculate the fractional moment of inertia of the crust $I_{crust}/I$ using the formula \cite{Fattoyev}
\begin{equation}
I_{crust} = \frac{16 \pi}{3} \frac{R_{core}^6 p_{PT}}{R_s} \left[ 1 - \frac{0.21}{(R/R_s-1)} \right]
\left[ 1 + \frac{48}{5}(R_{core}/R_s-1)(p_{PT}/\epsilon_{PT} + ...\right]~,
\end{equation}
where, the total moment of inertia of the star is
\begin{equation}
\frac{I}{M R^2} = \frac{0.21}{1-R_s/R}~.
\end{equation}
In the above expressions, $R_s = 2GM$ is the Schwarzschild radius of the star.

\begin{table*}[htbp]
   \caption{Effect of strong magnetic field on transition density and fractional moment of inertia of the crust for a neutron star of gravitational mass 1.4 M$_{\odot}$ with the SLy4 functional, in the isotropic TOV approximation. See text for details. }
\begin{tabular}{|c||c|c||c|c||c|}
\hline
 & & & & & \\
    $B_*$ & $\rho_1$ & $I^1_{crust}/I$ & $\rho_2$ & $I^2_{crust}/I$ &  $\Delta I_{crust}/I$  \\
    {} & (fm$^{-3}$) & {} & (fm$^{-3}$) & {} & {}\\
    \hline 
 \hline
 $0$ & 0.077 & 0.028 & 0.077 & 0.028 & 0 \\
 \hline
 $10^2$ & 0.076 & 0.028 & 0.076 & 0.028 & 0 \\
 \hline
 $10^3$ & 0.071 & 0.026 & 0.074 & 0.027 & 0.001  \\
  \hline
 $5 \times 10^3$ & 0.070 & 0.025 & 0.084 & 0.036 & 0.011 \\
  \hline
 $7 \times 10^3$ & 0.033 & 0.011 & 0.081 & 0.035 & 0.024 \\
  \hline
 $10^4$ & 0.041 & 0.015 & 0.074 & 0.030 & 0.015 \\
\hline
\end{tabular}
\label{tab:table_icrust}
\end{table*}

\newpage 
\subsection{Full numerical calculations}
\label{sec:lorene}

\begin{figure}[htbp]
  \begin{center}
      \includegraphics[width=.6\textwidth, angle=270]{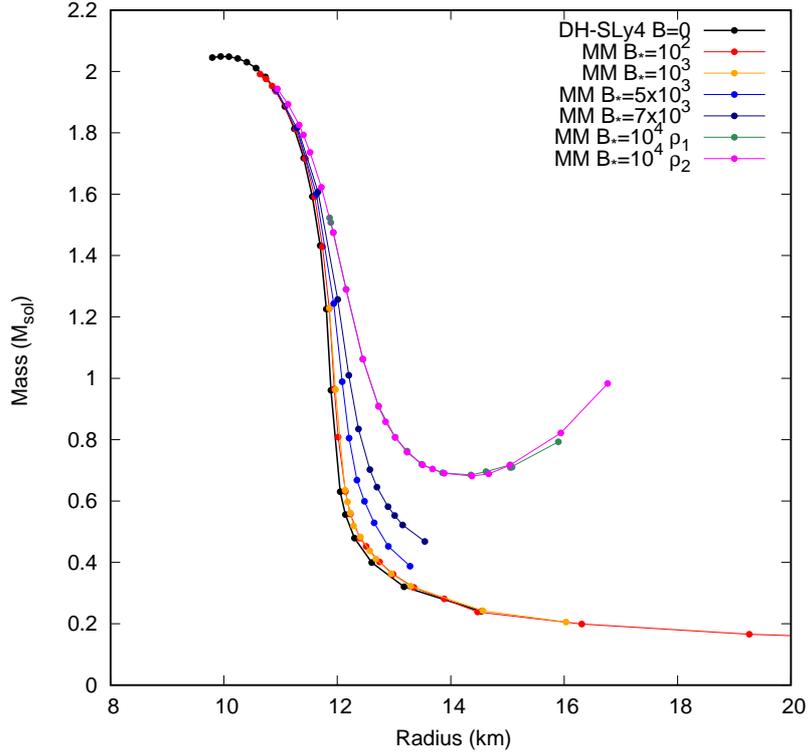}
    \caption{Circumferential radius vs gravitational mass of neutron stars endowed with strong magnetic fields computed using the complete MM (crust and core), and the full solution of the Einstein-Maxwell equations with LORENE,  for the SLy4 EoS.
    For comparison the Douchin Haensel SLy4 for B=0 is also plotted.}
    \label{fig:mrcompmm_sly4_allb}
  \end{center}
\end{figure}

As was already discussed, the isotropic TOV equations are not valid for strongly magnetized neutron stars for which the energy-momentum
tensor is anisotropic. There have been several suggestions in the literature to remedy the situation. Common procedures are to consider that the magnetic field contribution to the energy density and pressure is anisotropic in
the parallel and perpendicular directions to the magnetic field :
\begin{eqnarray}
\epsilon_{tot} &=& \epsilon_{had} + \epsilon_l + \epsilon_B \nonumber \\
p_{\parallel} &=& p_{had} + p_l - p_B \nonumber \\
p_{\perp} &=& p_{had} + p_l - {\cal M} B + p_B 
\end{eqnarray}
either neglecting or considering the contribution to the EoS from magnetization (the term ${\cal M} B$) \cite{MenezesLopes} .\\

Another alternative is the chaotic magnetic field formalism \cite{MenezesLopes}:
\begin{eqnarray}
\epsilon_{tot} &=& \epsilon_{had}  + \epsilon_l + \epsilon_B \nonumber \\
p_{tot} &=& p_{had} + p_l + p_B/3 
\end{eqnarray}

However, the ideal way to calculate the structure of strongly magnetized neutron stars is to solve the Einstein-Maxwell
and equilibrium solutions self-consistently with a magnetic field dependent EoS. 
We adopt this technique as we have at our disposal the numerical library LORENE \cite{Lorene} previously developed for studying the structure of strongly magnetized 
neutron stars \cite{Bocquet,Bonazzola1993,Chatterjee2014,Chatterjee2016}.
To this aim, we first construct total EoSs for neutron stars for the crust and the core obtained using the same MM. This is done using the compressible liquid drop model for the crust, with surface tension parameters optimized for each functional via a fit on nuclear masses \cite{Carreau}.
 The magnetic field dependence is only included in the core EoSs while assuming a non-magnetized crust, which is a reasonable approximation.\\

 We display the mass-radius relations corresponding to the SLy4 EoSs
for increasing magnetic fields in Fig.~\ref{fig:mrcompmm_sly4_allb}. It is evident from the figure that while neutron star mass-radius
relation for low fields $B_*=10^2, 10^3$ resembles that of the zero-field case, for higher magnetic fields $B_*=5 \times 10^3, 7 \times 10^3, 10^4$
the mass-radius relation departs strongly from the zero field case. As was discussed in \cite{Chatterjee2014,Chatterjee2016},
this is primarily due to the purely magnetic field contribution and not the effect of magnetic field on the EoS.
\\

\begin{figure}[htbp]
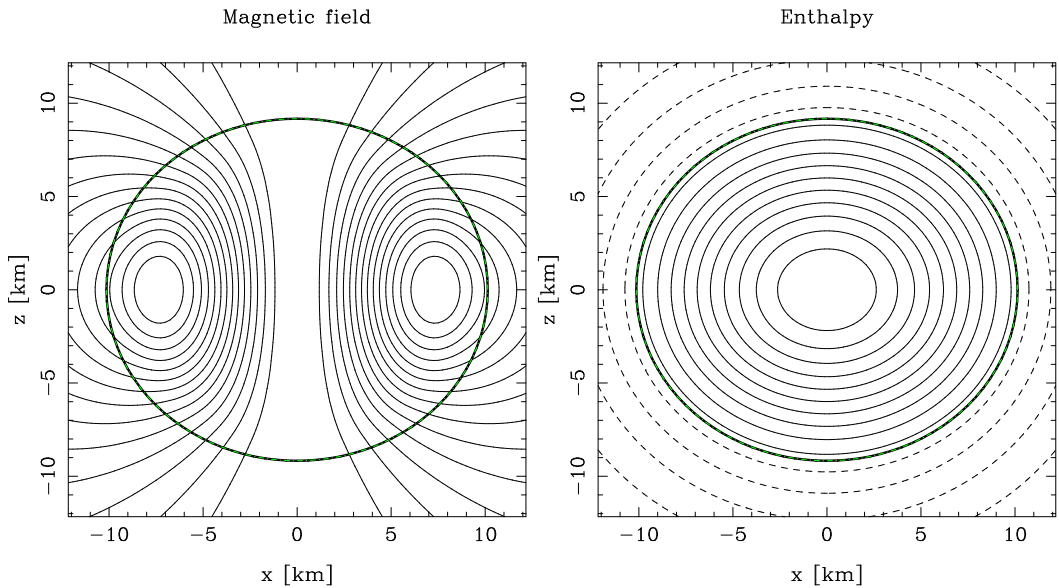

    \begin{center}
\includegraphics[width=0.45\textwidth, angle=270]{bfield.ps} 
\includegraphics[width=0.45\textwidth, angle=270]{enthalpy.ps} 
      \caption{ Magnetic field lines (left panel) and enthalpy isocontours (right panel) in the meridional plane (x, z), for the static star configuration,
with a gravitational mass of 1.4 $M_{\odot}$ and a polar magnetic field $B_*=10^4$. The stellar surface is depicted by the bold line. In the
right panel, solid lines represent positive enthalpy isocontours, dashed lines negative ones.
      }\label{fig:isocontours}
\end{center}
\end{figure}

\begin{figure}[htbp]
    \begin{center}
\includegraphics[width=0.45\textwidth, angle=270]{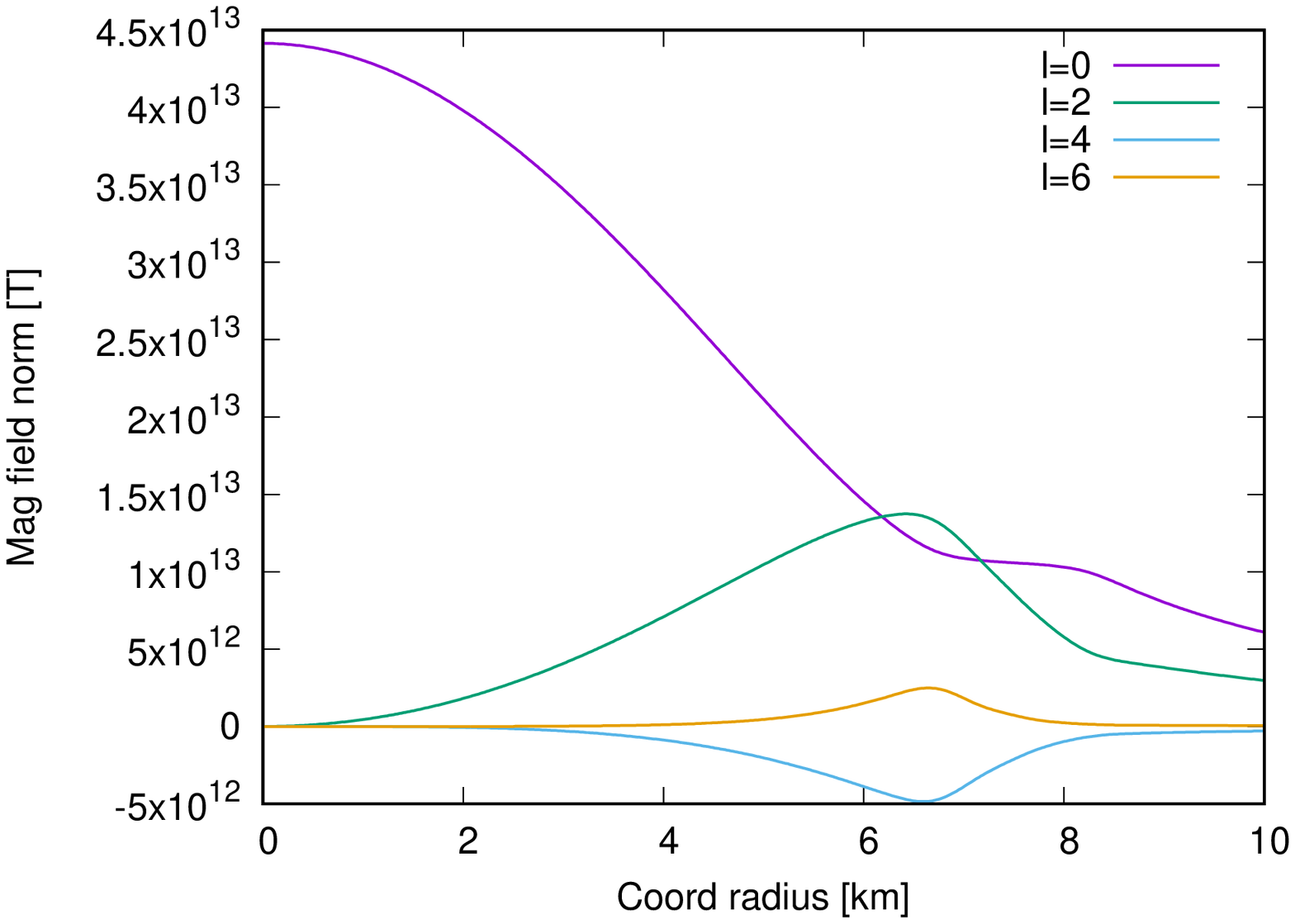} 
\includegraphics[width=0.45\textwidth, angle=270]{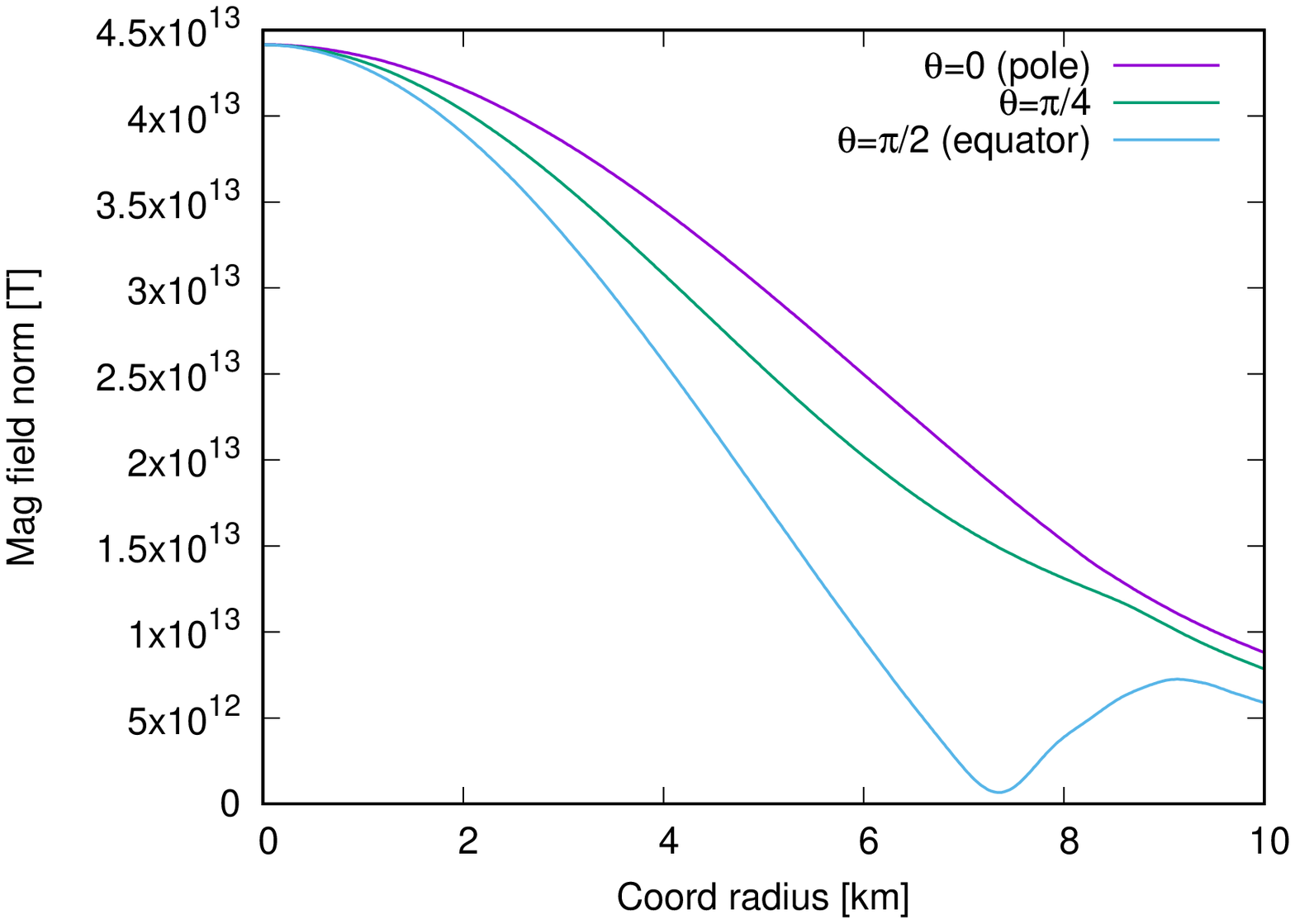} 
      \caption{ Magnetic field lines (left panel) and enthalpy isocontours (right panel) in the meridional plane (x, z), for the static star configuration,
with a gravitational mass of 1.4 $M_{\odot}$ and a polar magnetic field $B_*=10^4$.  
      }\label{fig:multipoles}
\end{center}
\end{figure}

One must also note that while for the isotropic TOV calculation one assumed a constant magnetic field,
the magnetic field for the full numerical calculation is generated via a current function and hence generates a profile, from the highest
central value decreasing towards the surface \cite{Chatterjee2018}. In Fig.~\ref{fig:isocontours} we display the magnetic field lines and enthalpy
isocontours of a neutron star of gravitational mass of 1.4 $M_{\odot}$ and a polar magnetic field $B_*=10^4$ (i.e. 4.414 $\times 10^{13}$ T). It is evident that the magnetic field
is far from constant, and the neutron star surface deviates from a spherical shape, as elaborated previously. 
As the neutron star radius as well as the magnetic field are direction dependent, we specify for this study the radius as 
the circumferential equatorial radius $R_{circ}$ (see \cite{Bonazzola1993}). In Fig.~\ref{fig:multipoles}, we display the 
radial profiles of the first four even multipoles ($l$ = 0, 2, 4, 6) of the magnetic field norm $b(r, \theta)$ computed
for the stellar model described in Fig.\ref{fig:isocontours}. From symmetry argument odd multipoles are all zero \cite{Chatterjee2018}.
We also plot the magnetic field norm as a function of the coordinate radius for different angular directions ($\theta = 0, \pi/4, \pi/2$).
It is clear that the magnetic field structure
in this case can neither be described by a constant amplitude nor a simple profile.
\\

Similar to Table~\ref{tab:table_lcrust}, we summarize in Table~\ref{tab:tablelorene_sly4} the effect of different magnetic field strengths
    on the transition densities and the neutron star structure computed within the full numerical structure calculation via LORENE. As before, for low magnetic fields 
    ($B_* \sim 10^{2}, 10^{3}$), there is a unique transition density.
    At higher fields, as discussed in Sec.~\ref{sec:resspi}, there are multiple transition density points, of which the
    first and the last crossings are denoted as $\rho_1$ and $\rho_2$ in the table.
    The difference in crust-core transition density leads to a difference in the corresponding total circumferential radii $R_{NS}$,
    circumferential core radii $R_{core}$ and crust thickness $l_{crust}$. The values of $l^1_{crust}$ and $l^2_{crust}$, 
    corresponding to $\rho_1$ and $\rho_2$, are also given in the table, along with their difference 
    $\Delta l_{crust} = l^2_{crust}- l^1_{crust}$. \\
    
    As compared to the previous Table ~\ref{tab:table_lcrust}, one can see here that although the total radii are comparable,
    the core radii and hence the crust thickness calculated numerically differ significantly from the isotropic TOV case. 
    This is due to the difference in the treatment of the magnetic field profile as well as the pure magnetic field contribution.
    In particular, the difference in the crust thickness $\Delta l_{crust}$ is significantly higher. Actually, both $l^1_{crust}$ and $l^2_{crust}$ are now of the order of 3 Km, considerably thicker than what was obtained with the TOV treatment for an isotopic star and also larger than the 2.4 Km, estimated in \cite{jorge2014}.\\

\begin{table*}[htbp]
   \caption{Effect of strong magnetic field on total circumferential radius, circumferential core radius and crust thickness for a neutron star of 
   gravitational mass 1.4 M$_{\odot}$ computed within a full numerical structure calculation with LORENE.  } 
\begin{tabular}{|c||c|c|c|c||c|c|c|c||c|}
\hline
    $B_*$ & $\rho_1$ & $R^1_{NS}$ & $R^1_{core}$ & $l^1_{crust}$ & $\rho_2$ & $R^2_{NS}$ & $R^2_{core}$ & $l^2_{crust}$ &  $\Delta l_{crust}$  \\
    {} & (fm$^{-3}$) & (km) & (km) & (km) & (fm$^{-3}$) & (km) & (km) & (km) & (km)\\
 \hline 
 \hline
 {}& {}& {}& {}&{}&{}&{}&{}&{}&{}\\ \hline 
 $10^2$ & 0.076 & 11.791 & 8.577 & 3.214 & 0.076 & 11.791 & 8.577 & 3.214 & 0 \\
 \hline
 $10^3$ & 0.074 & 11.782 & 8.595 & 3.187 & 0.074 & 11.782 & 8.595 & 3.187 & 0  \\
  \hline
 $5 \times 10^3$ & 0.070 & 11.836 & 8.705 & 3.131 & 0.084 & 11.841 & 8.448 & 3.393 & 0.262 \\
  \hline
 $7 \times 10^3$ & 0.033 & 11.900 & 8.585 & 3.315 & 0.081 & 11.896 & 8.320 & 3.576 & 0.261 \\
  \hline
 $10^4$ & 0.041 & 12.039 & 8.300 & 3.739 & 0.074 & 12.037 & 8.129 & 3.908 & 0.169\\
\hline
\end{tabular}
\label{tab:tablelorene_sly4}
\end{table*}

\subsection{Model dependence of the results} \label{sec:models}
\begin{figure}[htbp]
  \begin{center}
      \includegraphics[width=.6\textwidth, angle=270]{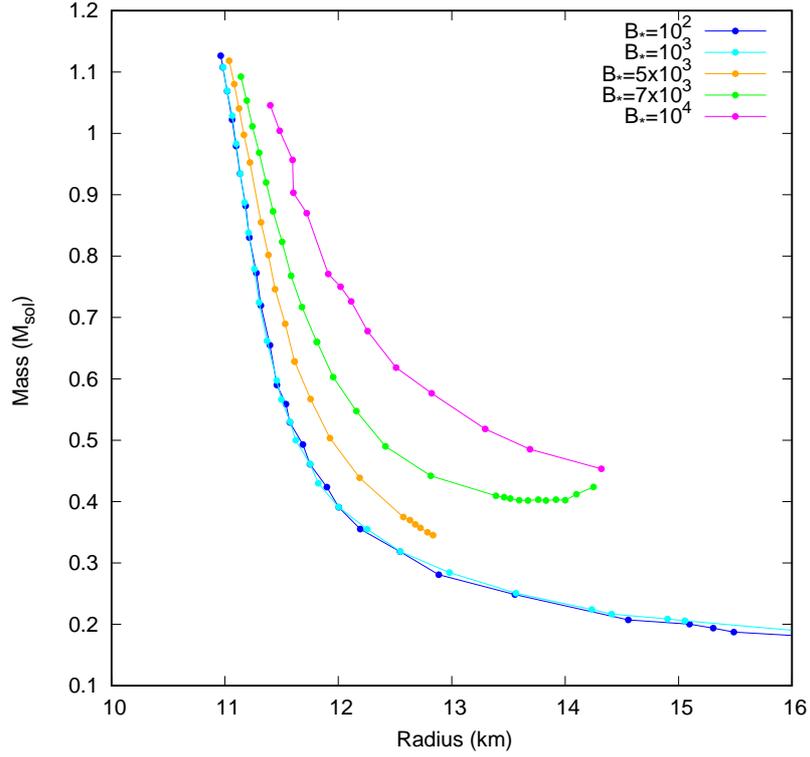}
    \caption{Circumferential radius vs gravitational mass of neutron stars endowed with strong magnetic fields computed using the MM for Bsk17 EoS.}
    \label{fig:mrcomp_bsk17_allb}
  \end{center}
\end{figure}

\begin{figure}[htbp]
  \begin{center}
      \includegraphics[width=.6\textwidth, angle=270]{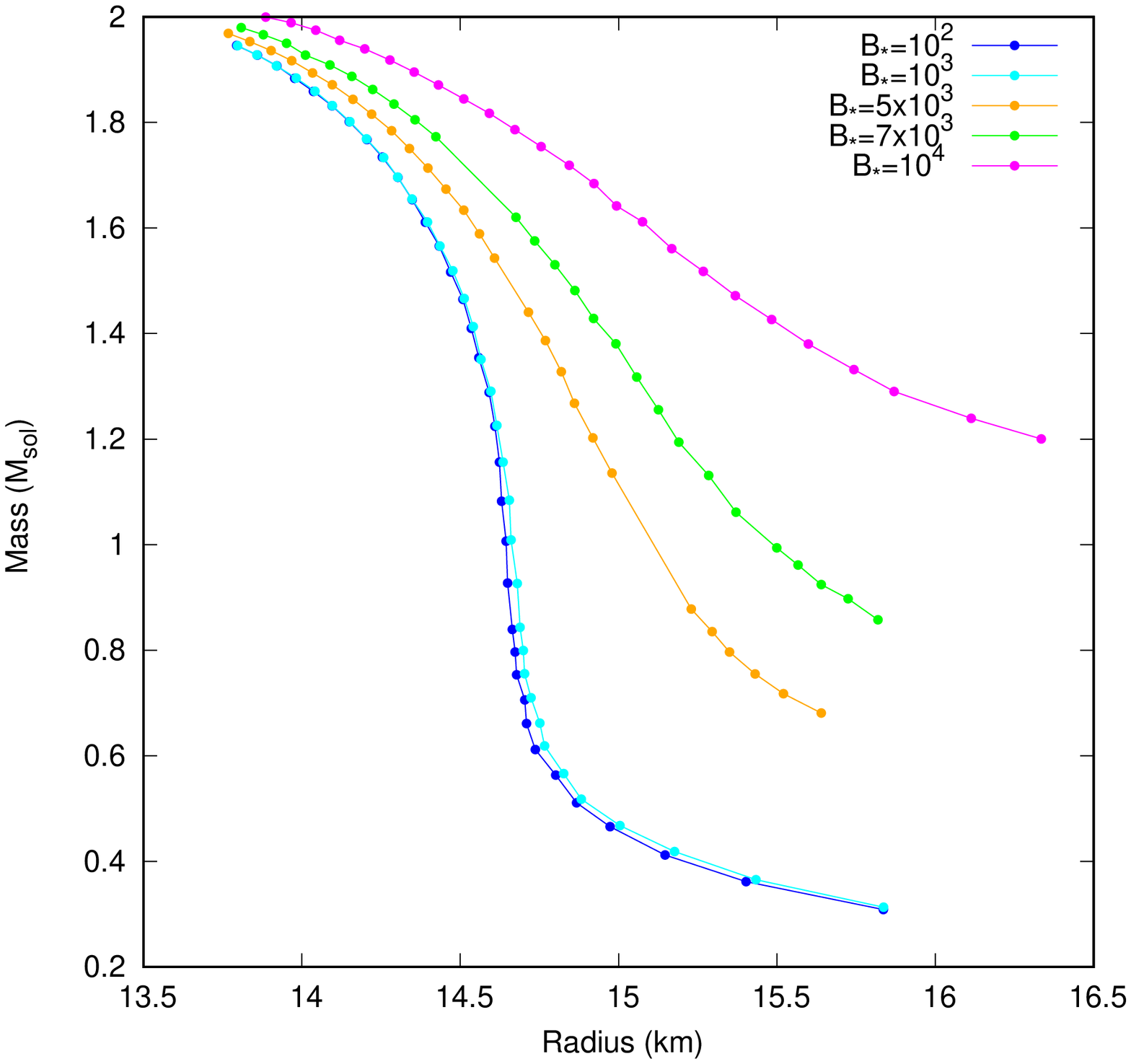}
    \caption{Circumferential radius vs gravitational mass of neutron stars endowed with strong magnetic fields computed using the MM for TM1 EoS.}
    \label{fig:mrcomp_tm1_allb}
  \end{center}
\end{figure}

In order to generalize the results obtained for any EoS, we perform the isotropic approximation and full numerical calculation for two other reference parameter sets, TM1 and Bsk17 (see Sec.\ref{sec:metamodel}). The corresponding mass-radius relations for various magnetic field values obtained numerically are given in the figures \ref{fig:mrcomp_bsk17_allb} and \ref{fig:mrcomp_tm1_allb} and
the main results are summarized in Tables \ref{tab:tablelorene_bsk17} and \ref{tab:tablelorene_tm1}.
\\

\begin{table*}[htbp]
   \caption{Effect of strong magnetic field on total circumferential radius, circumferential core radius and crust thickness for a neutron star of 
   gravitational mass 1.1 M$_{\odot}$ computed numerically, for the Bsk17 EoS. }
\begin{tabular}{|c||c|c|c|c||c|c|c|c||c|}
 \hline
    $B_*$ & $\rho_1$ & $R^1_{NS}$ & $R^1_{core}$ & $l^1_{crust}$ & $\rho_2$ & $R^2_{NS}$ & $R^2_{core}$ & $l^2_{crust}$ &  $\Delta l_{crust}$  \\
    {} & (fm$^{-3}$) & (km) & (km) & (km) & (fm$^{-3}$) & (km) & (km) & (km) & (km)\\
     {}& {}& {}& {}&{}&{}&{}&{}&{}&{}\\ \hline 
 \hline
 \hline
 $10^2$ & 0.090 & 10.995 & 8.660 & 2.335 & 0.090 & 10.995 & 8.660 & 2.335 & 0 \\
 \hline
 $10^3$ & 0.085 & 10.990 & 8.677 & 2.313 & 0.087 & 10.990 & 8.677 & 2.313 & 0  \\
  \hline
 $5 \times 10^3$ & 0.078 & 11.060 & 8.533 & 2.527 & 0.093 & 11.059 & 8.510 & 2.549 & 0.022 \\
  \hline
 $7 \times 10^3$ & 0.071 & 11.136 & 8.413 & 2.723 & 0.091 & 11.143 & 8.396 & 2.747 & 0.024 \\
  \hline
 $10^4$ & 0.048 & 11.244 & 8.179 & 3.065 & 0.115 & 11.276 & 8.090 & 3.186 & 0.121 \\
\hline
\end{tabular}
\label{tab:tablelorene_bsk17}
\end{table*}

\begin{table*}[htbp]
   \caption{Effect of strong magnetic field on total circumferential radius, circumferential core radius and crust thickness for a neutron star of gravitational mass 1.8 M$_{\odot}$ computed numerically, for the TM1 EoS.}
\begin{tabular}{|c||c|c|c|c||c|c|c|c||c|}
 \hline
    $B_*$ & $\rho_1$ & $R^1_{NS}$ & $R^1_{core}$ & $l^1_{crust}$ & $\rho_2$ & $R^2_{NS}$ & $R^2_{core}$ & $l^2_{crust}$ &  $\Delta l_{crust}$  \\
    {} & (fm$^{-3}$) & (km) & (km) & (km) & (fm$^{-3}$) & (km) & (km) & (km) & (km)\\
     {}& {}& {}& {}&{}&{}&{}&{}&{}&{}\\ \hline 
 \hline
 \hline
 $10^2$ & 0.066 & 14.153 & 10.777 & 3.376 & 0.066 & 14.153 & 10.777 & 3.376 & 0 \\
 \hline
 $10^3$ & 0.057 & 14.157 & 10.781 & 3.376 & 0.063 & 14.157 & 10.781 & 3.376 & 0  \\
  \hline
 $5 \times 10^3$ & 0.055 & 14.252 & 10.625 & 3.627 & 0.093 & 14.251 & 10.595 & 3.656 & 0.029 \\
  \hline
 $7 \times 10^3$ & 0.061 & 14.357 & 10.444 & 3.913 & 0.099 & 14.357 & 10.415 & 3.942 & 0.029 \\
  \hline
 $10^4$ & 0.038 & 14.594 & 10.119 & 4.475 & 0.102 & 14.856 & 10.055 & 4.531 & 0.056 \\
\hline
\end{tabular}
\label{tab:tablelorene_tm1}
\end{table*}

As it is already very well known, the stiffest model (TM1) produces the largest radius for a given mass. We can see from Figs.\ref{fig:mrcomp_bsk17_allb} and \ref{fig:mrcomp_tm1_allb} that this model dependence is preserved with the increase of the magnetic field. The width of the crust monotonically increases with the magnetic field for all studied models, and it seems to be very well correlated with the star radius, a stiffer EoS producing a thicker crust. 
Conversely, the region where successive homogeneous and inhomogeneous layers are expected \cite{cp1,cp2,cp3}, measured by  the variable $\Delta l_{crust}$, seems to decrease with increasing stiffness.

For the numerical calculations of the strongly magnetized neutron stars using LORENE, one may calculate the total moment of inertia of a slowly rotating neutron star from the total angular momentum, taking into account both contributions from rotation and magnetic field. However, within the numerical models (see \cite{Chatterjee2014} and references within), one cannot isolate the crustal moment of inertia in a gauge-independent way. Hence we refrain from performing a calculation of the fractional crustal moment of inertia as was done in Sec. \ref{sec:tov} for the TOV calculations.\\

\section{Final remarks}
\label{sec:conclusions}
 
In this work, we extended a recently developed  meta-modelling technique to study the crust-core phase transition properties and crustal thickness in the presence of strong magnetic fields.
It was found that the magnetic field severely modifies the structure of the phase transition region, leading to the a non-negligible difference 
in the density and pressure of the transition from that of the zero magnetic field case. 
As previously observed in refs.\cite{cp1,cp2,cp3}, the most spectacular effect of the magnetic field is the fact that the instable region with respect to density fluctuations is discontinuous if the magnetic field is sufficiently intense. As a consequence, beta equilibrated matter can correspond to a sequence of stable and unstable regions, as a function of the density. 
How this feature will reflect in the composition of the neutron star is not clear. It might be that the crust is simply more extended, or that in its inner part homogeneous and inhomogeneous layers could coexist, with much higher impurity factors than usually considered for non-magnetized star.

A sensitivity analysis to the most influential isovector empirical parameters ($L_{sym}$ and $K_{sym}$) has shown that the discontinuity in the instable region is strongly increasing with the $L_{sym}$ parameter, and the effect is amplified by the intensity of the magnetic field.

In order to study the effect of the change in crust-core transition properties on the neutron star structure, we adopted two different formalisms.
Firstly we calculated the mass-radius relation as well as the crust thickness using an isotropic TOV formalism, that has been commonly used in the 
literature. Next we performed a full self-consistent numerical computation of the neutron star structure. We showed that the results in the 
two cases vary considerably, and underlined the fact that a full numerical formalism is inevitable for the structure calculation of strongly
magnetized neutron stars.

\section*{Acknowledgments}
This work was partially supported by the NewCompStar COST action
MP1304, Capes(Brazil)/Cofecub (France), joint international
collaboration project number 853/15 and project INCT-FNA Proc. No. 464898/2014-5.
DC acknowledges the financial support from the CNRS/In2p3 through the Master project MAC and DPM was
partially supported by CNPq (Brazil) under grant 301155/2017-8.
DPM thanks the LPC-Caen for the hospitality.\\
The authors thank Sofija Anti\'c for providing useful data for comparison with the non-magnetic field case and Thomas Carreau for providing the crust EoS for the different functionals used in this work.
  

\end{document}